# A satellite orbit drift in binary near-Earth asteroids (66391) 1999 KW4 and (88710) 2001 SL9 — Indication of the BYORP effect

P. Scheirich [a,*], P. Pravec [a], P. Kušnirák [a], K. Hornoch [a], J. McMahon [b], D. J. Scheeres [b], D. Čapek [a], D. P. Pray [c], H. Kučáková [a,s,t], A. Galád [d], J. Vraštil [a], Yu. N. Krugly [e], N. Moskovitz [f], L. D. Avner [g], B. Skiff [f], R. S. McMillan [g], J. A. Larsen [g,h], M. J. Brucker [g], A. F. Tubbiolo [g], W. R. Cooney [i], J. Gross [i], D. Terrell [i,j], O. Burkhonov [k], K. E. Ergashev [k], Sh. A. Ehgamberdiev [k], P. Fatka [a], R. Durkee [ℓ], E. Lilly Schunova [p], R. Ya. Inasaridze [m,n], V. R. Ayvazian [m,n], G. Kapanadze [m,n], N. M. Gaftonyuk [o], J. A. Sanchez [p], V. Reddy [g], L. McGraw [q], M. S. Kelley [r], I. E. Molotov [u]

[a] *Astronomical Institute of the Czech Academy of Sciences, Fričova 1, CZ-25165 Ondřejov, Czech Republic*

[b] *Department of Aerospace Engineering Sciences, The University of Colorado at Boulder, Boulder, CO, USA*

[c] *Sugarloaf Mountain Observatory, South Deerfield, MA, USA*

[d] *Modra Observatory, Department of Astronomy, Physics of the Earth, and Meteorology, FMPI UK, Bratislava SK-84248, Slovakia*

[e] *Institute of Astronomy of Kharkiv National University, Sumska Str. 35, Kharkiv 61022, Ukraine*

[f] *Lowell Observatory, 1400 W Mars Hill Road, Flagstaff, AZ 86001, USA*

[g] *Lunar and Planetary Laboratory, University of Arizona, 1629 East University Boulevard, Tucson, AZ 85721, USA*

[h] *U.S. Naval Academy, Annapolis, MD, USA*

[i] *Sonoita Research Observatory, 77 Paint Trail, Sonoita, AZ 85637, USA*

[j] *Deptartment of Space Studies, Southwest Research Institute, Boulder, CO 80302, USA*

[k] *Ulugh Beg Astronomical Institute, Astronomicheskaya Street 33, 100052 Tashkent, Uzbekistan*

[ℓ] *Shed of Science South Observatory, Pontotoc, TX, USA*

[m] *Kharadze Abastumani Astrophysical Observatory, Ilya State University,*

*K. Cholokashvili Avenue 3/5, Tbilisi 0162, Georgia*

[n] *Samtskhe-Javakheti State University, Rustaveli Street 113, Akhaltsikhe 0080, Georgia*

[o] *Crimean Astrophysical Observatory of Russian Academy of Sciences, 298409 Nauchny, Ukraine*

[p] *Planetary Science Institute, 1700 E. Fort Lowell Road, Tucson, AZ 85719, USA*

[q] *Northern Arizona University, Flagstaff, AZ, USA*

[r] *Planetary Defense Coordination Office, NASA Headquarters, 300 E Street SW, Washington, DC 20546, USA*

[s] *Astronomical Institute, Faculty of Mathematics and Physics, Charles University, V Holešovičkách 2, 180 00, Prague 8, Czech Republic*

[t] *Research Centre for Theoretical Physics and Astrophysics, Institute of Physics, Silesian University in Opava, Bezručovo nám. 13, CZ-74601 Opava, Czech Republic*

[u] *Keldysh Institute of Applied Mathematics, RAS, Miusskaya sq. 4, Moscow 125047, Russia*

Proposed running head: Orbital drift in binary asteroids 1999 KW4 and 2001 SL9

Editorial correspondence to:
Peter Scheirich, Ph.D.
Astronomical Institute AS CR
Fričova 1
Ondřejov
CZ-25165
Czech Republic
Phone: 00420-323-620115
Fax: 00420-323-620263
E-mail address: petr.scheirich@gmail.com

**Abstract**

We obtained thorough photometric observations of two binary near-Earth asteroids (66391) Moshup = 1999 KW4 and (88710) 2001 SL9 taken from 2000 to 2019. We modeled the data and derived physical and dynamical properties of the binary systems. For (66391) 1999 KW4, we derived its mutual orbit's pole, semimajor axis and eccentricity that are in agreement with radar-derived values (Ostro et al. [2006]. Science, 314, 1276–1280). However, we found that the data are inconsistent with a constant orbital period and we obtained unique solution with a quadratic drift of the mean anomaly of the satellite of $-0.65 \pm 0.16$ deg/yr$^2$ (all quoted uncertainties correspond to $3\sigma$). This means that the semimajor axis of the mutual orbit of the components of this binary system, determined $a = 2.548 \pm 0.015$ km by Ostro et al. (2006), increases in time with a mean rate of $1.2 \pm 0.3$ cm/yr.

For (88710) 2001 SL9, we determined that the mutual orbit has a pole within 10° of $(L, B) = (302°, -73°)$ (ecliptic coordinates), and is close to circular (eccentricity $< 0.07$). The data for this system are also inconsistent with a constant orbital period and we obtained two solutions for the quadratic drift of the mean anomaly: $2.8 \pm 0.2$ and $5.2 \pm 0.2$ deg/yr$^2$, implying that the semimajor axis of the mutual orbit of the components (estimated $a \sim 1.6$ km) decreases in time with a mean rate of $-2.8 \pm 0.2$ or $-5.1 \pm 0.2$ cm/yr for the two solutions, respectively.

The expanding orbit of (66391) 1999 KW4 may be explained by mutual tides interplaying with binary YORP (BYORP) effect (McMahon, J., Scheeres, D. [2010]. Icarus 209, 494-509). However, a modeling of the BYORP drift using radar-derived shapes of the binary components predicted a much higher value of the orbital drift than the observed one. It suggests that either the radar-derived shape model of the secondary is inadequate for computing the BYORP effect, or the present theory of BYORP overestimates it. It is possible that the BYORP coefficient has instead an opposite sign than predicted; in that case, the system may be moving into an equilibrium between the BYORP and the tides.

In the case of (88710) 2001 SL9, the BYORP effect is the only known physical mechanism that can cause the inward drift of its mutual orbit.

Together with the binary (175706) 1996 FG3 which has a mean anomaly drift consistent with zero, implying a stable equilibrium between the BYORP effect and mutual body tides (Scheirich et al. [2015]. Icarus 245, 56-63), we now have three distinct cases of well observed binary asteroid systems with their long-term dynamical models inferred. They indicate a presence of all the three states of the mutual orbit evolution – equilibrium, expanding and contracting – in the population of near-Earth binary asteroids.

* Corresponding author. Fax: +420 323 620263.

*Email address:* petr.scheirich@gmail.com (P. Scheirich).

## 1 Introduction

Binary asteroids exhibit interesting mutual two-body dynamics driven by thermal emission from irregularly shaped components, but up to now there has appeared only one study constraining its limits based on direct measurements so far: Scheirich et al. (2015) found an upper limit on drift of the mutual orbit of the components of binary near-Earth asteroid (175706) 1996 FG3, that is consistent with the theory of Jacobson and Scheeres (2011) of that synchronous binary asteroids are in a state of stable equilibrium between binary YORP (BYORP) effect [1] and mutual body tides. In this paper, we present a comprehensive analysis of mutual orbit drifts in two well-observed binary near-Earth asteroids (NEAs).

The NEA (66391) Moshup = 1999 KW4 was discovered by Lincoln Near-Earth Asteroid Research in Socorro, New Mexico, on 1999 May 20. Its binary nature was revealed by Benner et al. (2001). We obtained thorough photometric observations for it in six apparitions from 2000 to 2019. Since the asteroid was named only recently and its original designation 1999 KW4 is well-known to the asteroid science community, we use it throughout this paper.

The NEA (88710) 2001 SL9 was discovered by Near-Earth Asteroid Tracking at Palomar on 2001 September 18. Its binary nature was revealed by Pravec et al. (2001). We obtained thorough photometric observations for it in five apparitions from 2001 to 2015.

Among binary NEAs observed so far, our photometric datasets for these three systems (together with 1996 FG3) are the longest coverages obtained, providing a unique opportunity to study evolution of the mutual orbits of components of small binary asteroids.

The structure of this paper is as follows. In Section 2, we present a model of the mutual orbit of the components of 1999 KW4 and 2001 SL9 constructed from our complete photometric datasets. Then in Sections 3 and 4, we summarize our results with already known parameters of the two binaries. In Section 5, we then discuss implications of the observed characteristics, especially on the BYORP theory, from the derived drifts of the mutual orbits.

[1] The BYORP effect is a secular change of the mutual orbit of the components of a binary asteroid due to the emission of thermal radiation from asymmetric shapes of the components. It was first theoretically proposed by Ćuk and Burns (2005).

## 2 Mutual orbit models of 1999 KW4 and 2001 SL9

### *2.1 Observational data*

Table 1
Observations of (66391) 1999 KW4

| Time span | No. of nights | Telescope | References |
|---|---|---|---|
| 2000-06-19.0 to 2000-06-29.0 | 5 | 0.65-m Ondřejov | P06 |
| 2001-06-03.2 to 2001-06-20.9 | 7 | 0.41-m River Oaks | P06 |
| | 4 | 0.65-m Ondřejov | P06 |
| 2016-06-07.9 to 2016-06-22.3 | 6 | 0.65-m Ondřejov | This work |
| | 6 | 0.5-m Sugarloaf Mountain | This work |
| 2017-06-01.8 to 2017-06-27.0 | 8 | 0.65-m Ondřejov | This work |
| | 6 | 0.5-m Sugarloaf Mountain | This work |
| 2018-06-05.9 to 2018-06-18.9 | 9 | 0.65-m Ondřejov | This work |
| 2019-05-31.1 to 2019-06-09.2 | 6 | 1.8-m Spacewatch II | This work |
| | 6 | 0.65-m Ondřejov | This work |
| | 5 | 0.5-m Sonoita | This work |
| | 3 | 0.5-m Sugarloaf Mountain | This work |
| | 3 | 0.5-m Shed of Science South | This work |

Reference: P06 (Pravec et al., 2006)

Table 2
Observations of (88710) 2001 SL9

| Time span | No. of nights | Telescope | References |
|---|---|---|---|
| 2001-10-10.9 to 2001-10-21.3 | 7 | 0.65-m Ondřejov | P06 |
| | 2 | 0.5-m Palmer Divide | P06 |
| 2012-09-11.9 to 2012-11-15.4 | 4 | 1.54-m La Silla | This work |
| | 4 | 1.5-m Maidanak | This work |
| 2013-10-12.0 to 2013-12-05.1 | 7 | 1.54-m La Silla | This work |
| | 2 | 0.7-m Abastumani | This work |
| | 1 | 1.0-m Simeiz | This work |
| 2014-10-18.0 to 2014-10-26.1 | 4 | 1.54-m La Silla | This work |
| 2015-07-09.2 to 2015-08-17.3 | 6 | 1.8-m Lowell | This work |
| | 3 | 2.2-m U. Hawaii | This work |

Reference: P06 (Pravec et al., 2006)

The data used in our analysis, obtained during six and five apparitions for 1999 KW4 and 2001 SL9, respectively, are summarized in Tables 1 and 2. The references and descriptions of observational procedures of the individual observatories are summarized in Table 3.

The data were reduced using the standard technique described in Pravec et al.

Table 3
Observational stations

| Telescope | Observatory | References for observational and reduction procedures |
|---|---|---|
| 2.2-m U. Hawaii | Mauna Kea, Hawaii | 1 |
| 1.8-m Lowell | Lowell Observatory, Arizona | 2 |
| 1.8-m Spacewatch II | Spacewatch, Arizona | M07, L20 |
| 1.54-m La Silla | La Silla, European Southern Observatory, Chile | P14 |
| 1.5-m Maidanak | Maidanak Astronomical Observatory, Uzbekistan | P19 |
| 1.0-m Simeiz | Simeiz, Crimea | 3 |
| 0.7-m Abastumani | Abastumani, Georgia | K16, P19 |
| 0.65-m Ondřejov | Ondřejov, Czech Republic | P06 |
| 0.5-m Sugarloaf Mountain | Sugarloaf Mountain Observatory, Massachusetts | V17 |
| 0.5-m Sonoita | Sonoita Research Observatory, Arizona | C15 |
| 0.5-m Palmer Divide | Palmer Divide Observatory, Colorado | P06 |
| 0.5-m Shed of Science South | Shed of Science South Observatory, Texas | 4 |
| 0.41-m River Oaks | River Oaks Observatory, Texas | P06 |

References: 1: The observations were made in the Cousins R filter. Standard procedure of image reduction included dark removal and flatfield correction. 2: The observations were reduced using the same procedure as the observations from the 1.54-m La Silla, see Pravec et al. (2014) for details. 3: The observations were carried with a 1-m Ritchey-Chrétien telescope at Simeiz Department of the Crimean Astrophysical Observatory using camera FLI PL09000. The observations were made in the Johnson-Cousins photometric system. Standard procedure of image reduction included dark removal and flatfield correction. The aperture photometry was done with the AstPhot package described in Mottola et al. (1995). The differential lightcurves were calculated with respect to an ensemble of comparison stars by the method described in Erikson et al. (2000) and Krugly (2004). 4: The Shed of Science South utilizes a 0.5m Corrected Dall Kirkham telescope operating at a focal ratio of f4.5 and a pixel scale of 1.24 arc seconds per pixel using an SBIG ST10XME. Flat, dark, and bias images were applied using MaximDL and photometry was done using MPO Canopus. All images were unfiltered. C15 (Cooney et al., 2015), K16 (Krugly et al., 2016), L20 (Larsen, J. A., et al. 2020. In preparation.), M07 (McMillan et al., 2007), P06 (Pravec et al., 2006), P14 (Pravec et al., 2014), P19 (Pravec et al., 2019), V17 (Vokrouhlický et al., 2017).

(2006). By fitting a two-period Fourier series to data points outside mutual (occultation or eclipse) events, the rotational lightcurves of the primary (short-period) and the secondary (long-period), which are additive in linear flux units, were separated. The long-period component containing the mutual events and the secondary rotation lightcurve is then used for subsequent numerical modeling.

## *2.2 Numerical model*

We constructed models of the two binary asteroids using the technique of Scheirich and Pravec (2009) that was further developed in Scheirich et al. (2015). In following, we outline the basic points of the method, but we refer the reader to the 2009 and 2015 papers for details of the technique.

The shapes of the binary asteroid components were represented with ellipsoids, orbiting each other on a Keplerian orbit with apsidal precession and allowing for a quadratic drift in mean anomaly. The primary was modeled as an oblate spheroid, with its spin axis assumed to be normal to the mutual orbital plane of the components (i.e., assuming zero inclination of the mutual orbit). The shape of the secondary was modeled as a prolate spheroid in synchronous rotation, with its long axis aligned with the centers of the two bodies (i.e., assuming zero libration). The shapes were approximated with 1016 and 252 triangular facets for the primary and the secondary, respectively. The components were assumed to have the same albedo. The brightness of the system as seen by the observer was computed as a sum of contributions from all visible facets using a ray-tracing code that checks which facets are occulted by or are in shadow from the other body. A combination of Lommel-Seeliger and Lambert scattering laws was used (see, e.g., Kaasalainen et al., 2002).

The quadratic drift in mean anomaly, $\Delta M_d$, was fitted as an independent parameter. It is the coefficient in the second term of the expansion of the time-variable mean anomaly:

$$M(t) = M(t_0) + n(t - t_0) + \Delta M_d (t - t_0)^2, \tag{1}$$

where

$$\Delta M_d = \frac{1}{2}\dot{n}, \tag{2}$$

where $n$ is the mean motion, $t$ is the time, and $t_0$ is the epoch. $\Delta M_d$ was stepped from $-15$ to $+15$ deg/yr$^2$ in the case of 1999 KW4 and from $-9$ to $+39$ deg/yr$^2$ in the case of 2001 SL9 and all other parameters were fitted at each step.[2] The steps in $\Delta M_d$ were 0.005 deg/yr$^2$ and 0.01 deg/yr$^2$ in the case of 1999 KW4 and 2001 SL9, respectively.

To reduce the complexity of the model, we estimated upper limits on the eccentricity of the mutual orbits by fitting the data from the best covered apparitions: the 2001 apparition for 1999 KW4 and the 2013 apparition for 2001 SL9. The model includes a precession of the line of apsides. The pericenter drift rate depends on the mutual orbital period, the mutual semimajor axis and on the polar flattening of the primary (see Murray and Dermott, 1999, Eq. (6.249)). While we know the first two parameters quite well, the polar flattenings are poorly constrained from the data (see Tables 4 and 5). We therefore fit the drift rate as an independent parameter. Its initial values were stepped in a range from 0 to 25°/day. This range reflects the uncertainties of the three parameters determining the drift rate.

[2] $\Delta M_d$ of 2001 SL9 was sampled on the larger interval because in our initial modeling runs, there appeared possible solutions at high positive $\Delta M_d$ values. Therefore, we expanded the interval in order to examine them; there turned out to be no significant solution at high $\Delta M_d$ finally.

Since we found that the upper limits on eccentricity were low, in further modeling of the data from all apparitions together, we set the eccentricity equal to zero for simplicity and efficiency. This assumption had a negligible effect on the accuracy of other derived parameters of the models.

Across all observations, we found a unique solution for the system parameters except for an ambiguity in the quadratic drift in mean anomaly and the orbital period of 2001 SL9, see Tables 4 and 5. We describe and discuss these parameters in Sections 3 and 4. Plots of the RMS residuals (root mean square of observed magnitudes minus the values calculated from the model) vs $\Delta M_d$ are shown in Figs. 1 and 3. In order to save computing time, the plots were constructed using spherical shapes of both components. However, neighborhoods of local minima were then revisited using ellipsoidal shapes in order to improve the fit.

For 1999 KW4, the RMS residuals of the two best local minima obtained using the spherical shapes (with $\Delta M_d$ of $-0.65$ and $-1.3$ deg/yr$^2$) were 0.0307 and 0.0320 mag, respectively. The fits improved to 0.0251 and 0.0266 mag using the elipsoidal shapes. The fit is significanly poorer for the latter solution. The former solution provides a satisfactory fit to the data and it is accepted as real solution for the binary asteroid parameters.

For 2001 SL9, the RMS residuals of the five best local minima obtained using the spherical shapes (with $\Delta M_d$ of 2.8, 5.2, 7.6, 4.0 and 0.5 deg/yr$^2$) were 0.0238. 0.0238, 0.0245, 0.0246 and 0.0248 mag, respectively. The fits improved to 0.0236, 0.0236, 0.0243, 0.0245 and 0.0245 mag using the elipsoidal shapes; the marginal improvement is due to that the secondary of 2001 SL9 is not prominently elongated. The first two solutions provide satisfactory fit to the data; one of them is a real solution for the binary asteroid parameters, but we cannot resolve this ambiguity with the available data. The other three solutions with the higher RMS residuals provide significantly poorer fits to the data and they do not appear real.

Figures 6 and 11 show the quadratic drift in the mean anomaly, $\Delta M$, which was computed as follows. We generated a synthetic lightcurve using model with parameters from the best-fit solution except $\Delta M_d$, which was fixed at zero. Then, for each apparition separately, we fitted the mean anomaly of the model in order to obtain the best match between its synthetic lightcurve and the observed data. $\Delta M$ is a difference between the mean anomaly of the original model and the fitted one.

Examples of the long-period component data together with the synthetic lightcurves of the best-fit solutions are presented in Figs. 5, 9 and 10. Uncertainty areas of the orbital poles are shown in Figs. 7 and 12.

We estimated realistic uncertainties of the fitted parameters using the procedure described in Scheirich and Pravec (2009). For each parameter, we obtained its admissible range that corresponds to a 3-$\sigma$ uncertainty.

## 2.3 Statistical significance of the solutions for $\Delta M_d$

The residuals of the model fitted to the observational data do not obey the Gaussian statistics because of systematic errors resulting from model simplifications. In particular, the residuals of nearby measurements appear correlated.

To eliminate the effect we adopted following strategy for evaluating the statistical significance, based on the $\chi^2$ test, of the solutions for $\Delta M_d$.

We choose a correlation time $d$ and for each data point $(i)$ we calculated how many other data points, $K_i$, are within $\pm d/2$ from the given point. We then applied a weight of $1/K_i$ to the data point in the $\chi^2$ sum. We also calculated an effective number of data points as $N_{\rm eff} = \sum_{i=1}^{N} 1/K_i$, where $N$ is the total number of data points. For normalized $\chi^2$ we than have $\chi^2 = 1/(N_{\rm eff} - M) \sum_{i=1}^{N} (O-C)_i^2/(\sigma_i^2 K_i)$, where $M$ is the number of fitted parameters of the model and $\sigma_i$ is a standard deviation of the $i$th point. As the residuals are predominated by model rather than observational uncertainties, we assign each data points the same standard deviation $\sigma_i = \sigma$, where $\sigma$ is the RMS residual of the best fit solution.

The procedure described above is equivalent to reducing the number of data points to one in each time interval with the length $d$ (i.e., to reducing the total number of point to $N_{\rm eff}$) and assigning $(O-C)^2$ of this point to be a mean of $(O_i - C_i)^2$ of all the points within the interval. However, our approach has the advantage that it does not depend on a particular realisation of dividing the observing time span into intervals of length $d$.

We choose the correlation time $d$ to be equal to 1/2 of the mean duration of a descending/ascending branch of the secondary mutual event, i.e., the mean time between the first and the second or between the third and the fourth contact. For 1999 KW4 and 2001 SL9, they are $d = 0.30$ h and 0.15 h, respectively. We also tested them with $d$ twice as long, i.e., equal to the full mean duration of the secondary event branch, but we found it to be inadequate as the longer correlation time resulted in a substantial loss of information by a too big reduction of the data points.

We note that the mutual orbit model fit is sensitive only to data points in mutual events and their closest neighborhood. Therefore we limited the above analysis only to such data points; points further outside events were not used as they contain information on secondary's rotation, but they do not effectively contribute to determination of the mutual orbit.

Plots of the normalized $\chi^2$ vs $\Delta M_d$ for values of $\Delta M_d$ close to the best-fit solutions are shown in Figs. 2 and 4 for 1999 KW4 and 2001 SL9, respectively. The plots were constructed using elipsoidal shapes of the components. p-values of the $\chi^2$ test (see, e.g., Feigelson and Babu 2012), corresponding to 1, 2 and $3\sigma$ interval of the $\chi^2$ distribution with $(N_{\rm eff} - M)$ degrees of freedom, are also shown.

## 3 Parameters of (66391) 1999 KW4

In this section, we summarize the best-fit model parameters of the binary system (66391) 1999 KW4 and overview previous publications. The parameters are listed in Table 4.

In the first part of the table, we present data derived from optical and spectroscopic observations of the system. $H_V$ and $G$ are the mean absolute magnitude and the phase parameter of the $H$–$G$ phase relation (Bowell et al., 1989). Using $H_V$ and effective diameter of the whole system ($D_{\rm eff} \equiv (D^2_{1,\rm C} + D^2_{2,\rm C})^{1/2}$) at the mean observed aspect of 27° (see below), we derived the visual geometric albedo $p_{\rm V}$. We note that our value is in agreement with $p_{\rm V} = 0.19 \pm 0.05$ derived by Devogèle et al. (2019) from their polarimetric observations. We also observed 1999 KW4 in near-infrared spectral range and classified it as a Q type asteroid (see Appendix A.).

In the next two parts of Table 4, we give parameters for the components of the binary. The indices 1 and 2 refer to the primary and the secondary, respectively.

$D_{i,\rm C}$ is the cross-section equivalent diameter, i.e., the diameter of a sphere with the same cross section, of the $i$-th component at the observed aspect. Since the aspect changed with time, the given value is an average over all lightcurve sessions. To quantify the mean aspect we used an asterocentric latitude of the Phase Angle Bisector (PAB), which is the mean direction between the heliocentric and geocentric directions to the asteroid. As discussed in Harris et al. (1984), this is an approximation for the effective viewing direction of an asteroid observed at non-zero solar phase. The average absolute value of the asterocentric latitude of the PAB (computed using the nominal pole of the mutual orbit; we assume that the spin poles of both components are the same as the orbit pole) was 27°.

$D_{i,\rm V}$ is the volume equivalent diameter, i.e., the diameter of a sphere with the same volume, of the $i$-th component. $D_{2,\rm C}/D_{1,\rm C}$ is the ratio between the cross-section equivalent diameters of the components. $P_i$ is the rotational period of the $i$-th component.

An analysis of the best subset of data for the secondary rotation taken from 2018-06-07.9 to -11.0 gave a formal best-fit estimate for the secondary rotation period of $17.53 \pm 0.12$ h ($3\sigma$; this includes also a synodic-sidereal difference uncertainty). This agrees with the mutual orbit period, within the error bar. Considering that all the observed secondary lightcurve minima coincide with or lie close to the mutual events —small differences may be due to a phase effect or secondary libration—, it is very likely that the secondary is in synchronous rotation. We therefore assume that $P_2$ is equal to the orbital period (see Table 4).

$(A_1B_1)^{1/2}/C_1$ is a ratio between the mean equatorial and the polar axes of the primary. $A_i/B_i$ is a ratio between the equatorial axes of the $i$-th component (equatorial elongation). $\rho_1 = \rho_2$ are the bulk densities of the two components, which we assumed to be the same in our modeling.

Table 4
Properties of binary asteroid (66391) 1999 KW4.

| Parameter | Value | Unc. | Reference |
|---|---|---|---|
| Whole system: | | | |
| $H_V$ | $16.74 \pm 0.22$ | $1\sigma$ | This work |
| $G$ | $(0.24 \pm 0.11)^a$ | $1\sigma$ | This work |
| $p_\mathrm{V}$ | $0.162 \pm 0.034$ | $1\sigma$ | This work |
| Taxon. class | Q | | This work |
| Primary: | | | |
| $D_{1,\mathrm{C}}$ (km) | $1.367 \pm 0.041^b$ | $1\sigma$ | From O06 |
| $D_{1,\mathrm{V}}$ (km) | $1.317 \pm 0.040$ | $1\sigma$ | O06 |
| $P_1$ (h) | $2.7645 \pm 0.0003$ | $1\sigma$ | O06 |
| $(A_1B_1)^{1/2}/C_1$ | $\leq 1.6^c$ / $1.17 \pm 0.15$ | $3\sigma$ | This work / O06 |
| $A_1/B_1$ | $1.04 \pm 0.04$ | $1\sigma$ | O06 |
| $\rho_1$ (g cm$^{-3}$) | $1.3^{+0.7}_{-0.4}$ / $1.97 \pm 0.72$ | $3\sigma$ | This work / O06 |
| Secondary: | | | |
| $D_{2,\mathrm{C}}/D_{1,\mathrm{C}}$ | $0.42 \pm 0.03^d$ | $3\sigma$ | This work |
| $D_{2,\mathrm{C}}$ (km) | $0.574 \pm 0.066$ | $3\sigma$ | This work |
| $D_{2,\mathrm{V}}$ (km) | $(0.59 \pm 0.04)^e$ | $1\sigma$ | This work |
| $P_2$ (h) | $(17.46)^f$ | | This work |
| $A_2/B_2$ | $1.3^{+0.3}_{-0.1}$ | $3\sigma$ | This work |
| Mutual orbit: | | | |
| $a/(A_1B_1)^{1/2}$ | $1.7 \pm 0.2$ | $3\sigma$ | This work |
| $a$ (km) | $2.548 \pm 0.015$ | $1\sigma$ | O06 |
| $(L_\mathrm{P}, B_\mathrm{P})$ (deg.) | $(329.6, -62.3) \pm (12 \times 4)^g$ | $3\sigma$ | This work |
| $P_\mathrm{orb}$ (h) | $17.45763 \pm 0.00004^h$ | $3\sigma$ | This work |
| $L_0$ (deg.) | $40 \pm 5^h$ | $3\sigma$ | This work |
| $e$ | $\leq 0.006$ | $3\sigma$ | O06 |
| $\Delta M_d$ (deg/yr$^2$) | $-0.65 \pm 0.16$ | $3\sigma$ | This work |
| $\dot{P}_\mathrm{orb}$ (h/yr) | $0.00013 \pm 0.00003$ | $3\sigma$ | This work |
| $\dot{a}$ (cm/yr) | $1.2 \pm 0.3$ | $3\sigma$ | This work |

References: O06 (Ostro et al., 2006)

[a] The range of high solar phase angles covered by the observations did not allow to determine the $G$ parameter. We assumed the mean $G$ value for S-complex asteroids (Warner et al., 2009).

[b] Derived from the primary shape model by O06 and for the average observed aspect. See text for details.

[c] The formal best-fit value is 1.1.

[d] This is a ratio of the cross-section equivalent diameters for the average observed aspect of 27°. See text for details.

[e] Derived using the shape model of the secondary from O06 rescaled by 130%. See text for details.

[f] The secondary appears to be in synchronous rotation. See text for details.

[g] These are the semiaxes of the uncertainty area; see its actual shape in Fig. 7.

[h] The $P_\mathrm{orb}$ and $L_0$ values are for epoch JD 2455305.0 (asterocentric time, i.e., light-time corrected), for which $P_\mathrm{orb}$ and $\Delta M_d$ do not correlate.

Most of the reported quantities have been derived as the parameters of our model described in Section 2.2 fitted to our observational data. Some values were taken or derived using data from other sources as we describe in following.

The cross-section and volume equivalent diameters of the primary were derived using the shape model of the primary from Ostro et al. (2006). Assuming its spin pole is the same as the mutual orbit pole (see below), we computed its rotationally averaged cross-section for each lightcurve session and present the mean value over all sessions. Its $1\sigma$ uncertainty was computed using the uncertainties of the dimensions of the primary from Ostro et al. (2006).

$D_{1,\mathrm{V}}$ was taken from Table 2 of Ostro et al. (2006).

$D_{2,\mathrm{V}}$ was derived using the shape model of the secondary by Ostro et al. (2006), rescaled to 130% of its original size to match mutual events' depths from our data (see below). Its $1\sigma$ uncertainty is a formal value taken from Table 2 of Ostro et al. (2006), but the real uncertainty may be higher because of uncertainties of the secondary radar shape model (Lance Benner, personal communication).

In the last part of Table 4, we summarize the parameters of the mutual orbit of the binary components. $a$ is the semimajor axis, $L_\mathrm{P}, B_\mathrm{P}$ are the ecliptic coordinates of the orbital pole in the equinox J2000, $L_0$ is the length of the secondary (i.e., the sum of angular distance from the ascending node and the length of the ascending node) for epoch JD 2455305.0, $e$ is the orbit eccentricity (only its upper limit was determined), and $\Delta M_d$ is the quadratic drift in mean anomaly. Since the orbital period $P_\mathrm{orb}$ changes in time, the value presented in Table 4 is valid for epoch JD 2455305.0. For this epoch, which is approximately the mean time of all observed events, a correlation between $P_\mathrm{orb}$ and $\Delta M_d$ is zero. We also give the time derivatives of the orbital period and the semimajor axis, derived from $\Delta M_d$.

Although the orbit of 1999 KW4 crosses those of Earth, Venus and Mercury, according to JPL HORIZONS system the asteroid experienced only four close approaches to Earth between 2000 and 2019. The approaches took place in May 2001, May 2002, May 2018 and May 2019 at distances of 0.032, 0.089, 0.078 and 0.035 AU, respectively. Since the observed mutual orbit period increase is based on the observations at six effective epochs (apparitions), we can rule out planetary-tug effects as a potential cause for the increase.

The uncertainty area of the orbit pole is shown in Fig. 7. The size of the area shrinks with increasing the flattening of the primary $(A_1B_1)^{1/2}/C_1$. To demonstrate the effect, we constrained the orbit pole uncertainties using three fixed values of the flattening (1.0, 1.2 and 1.4) and plotted the respective areas in the figure.

The uncertainties of the mutual semimajor axis and flattening of the primary are the main sources of the uncertainty of the bulk density of the system. In addition to that, the uncertainties of the two parameters are not independent. We therefore stepped $a$ and $(A_1B_1)^{1/2}/C_1$ on a grid (while all other parameters were fitted at each step) to obtain an uncertainty area of both parameters together. The area is shown in Fig. 8 with values of the bulk density for each combination of the parameters indicated.

The mutual orbit and shapes of the binary asteroid components of 1999 KW4 were modeled by Ostro et al. (2006) with radar observations taken in 2001. They report the size of the primary to be close to a tri-axial ellipsoid with axes 1417 $\times$ 1361 $\times$ 1183 m ($1\sigma$ uncertainties of $\pm$ 3%), and the secondary to be a tri-axial ellipsoid with axes 595 $\times$ 450 $\times$ 343 m ($1\sigma$ uncertainties of $\pm$ 5%). The dimensions given are extents of dynamically equivalent equal-volume ellipsoid (DEEVE; a homogeneous ellipsoid having the same moments of inertia and volume as the shape model).

They also found the parameters of the mutual orbit to be as follows: orbital period $P_{\rm orb} = 17.422 \pm 0.036$ h, semimajor axis $a = 2548 \pm 15$ m, eccentricity $e = 0.0004 \pm 0.0019$, pole direction in ecliptic coordinates: $L_{\rm P} = 325.8 \pm 3.5$ deg, $B_{\rm P} = -61.8 \pm 1.2$ deg (uncertainties correspond to $1\sigma$).

To compare our results with the values from Ostro et al. (2006), we computed $(A_1 B_1)^{1/2}/C_1$ and $a/(A_1 B_1)^{1/2}$ using their DEEVE for the primary and their semi-major axis of the mutual orbit. The result is plotted as a solid point in Fig. 8 with $1\sigma$ error bars.

There is one significant discrepancy between our results and those by Ostro et al. (2006): We obtained a significantly larger secondary-to-primary size ratio. To compare their result with ours, we computed a mean (rotationally averaged) cross-section ratio from the component shapes by Ostro et al. (2006): $(D_{2,\rm C}/D_{1,\rm C})_{\rm radar} = 0.34 \pm 0.02$ ($1\sigma$) at the same mean aspect as our observations (asterocentric latitude of the Phase Angle Bisector, $B_{\rm PAB} = 27^\circ$). The value is significantly lower than our $D_{2,\rm C}/D_{1,\rm C} = 0.42 \pm 0.03$ ($3\sigma$).

To look more into the discrepancy between the secondary-to-primary size ratios by Ostro et al. (2006) and by us, we performed following test. Using the shape models of both components from Ostro et al. and the orbital parameters from Table 4, we generated a synthetic long-period component of the lightcurve. We then increased the size of the secondary until the depths of the secondary events (occultations and eclipses of the secondary) matched the observed event depths. We obtained a match when we increased the secondary axes by Ostro et al. (2006) to 130% of their original values. This is even slightly greater than $0.42/0.34 \doteq 124\%$ because in this test a more realistic scattering model (a combination of Lommel-Seeliger and Lambert scattering) was used for calculating the synthetic lightcurve, which models the scattering from non-spherical component shapes at the high solar phases and it is more precise than simply comparing the estimated mean cross-sections above. We note that replacing the parameters of the mutual orbit with those derived by Ostro et al. did not change the result.

We discussed this issue with Lance Benner and we received following information: "The dimensions of the secondary might be underestimated by Ostro et al. (2006) because the radar images were obtained at relatively coarse range and Doppler resolutions and at modest signal-to-noise ratios. Consequently, it is plausible that the trailing edge of the secondary in the radar images were less than would be detected if the SNRs were substantially higher." (Lance Benner, personal communication.)

## 4 Parameters of (88710) 2001 SL9

In this section, we summarize the best-fit model parameters of the binary system (88710) 2001 SL9 and overview previous publications. The parameters are listed in Table 5.

The notation of the values in the table and their uncertainties are the same as in Table 4 (see Section 3).

The average absolute value of the asterocentric latitude of the PAB (computed using the nominal pole of the mutual orbit, assumed to be the rotation pole of both components) was 11°; we observed the asteroid close to equator-on.

Three works were published reporting spectroscopic observations of 2001 SL9 in the visual and near-infrared spectral range: Lazzarin et al. (2004, 2005) and Pajuelo et al. (2018). Based on moderate slope and broad $1\mu$m and $2\mu$m absorbtion bands, Lazzarin et al. (2004) and (2005) classified 2001 SL9 as an Sr and Q type, respectively. Pajuelo et al. found that the taxonomic types that fit their NIR spectrum are Sr, S and Sq, with Sr being the best fit.

From the measured $H_V$ and assuming the mean albedo $p_V = 0.20 \pm 0.05$ for S-complex asteroids (Pravec et al., 2012), we estimated the effective diameter of the system $D_{\rm eff}$ at the observed (near equator-on) aspect.

A rotational state of the secondary is particularly important for the interpretations we present in Section 5. However, as the amplitude of the secondary rotation lightcurve is very low, we could not derive its rotation period from the available data. It appears that the secondary is nearly spheroidal with low equatorial elongation.

Pravec et al. (2016) showed that asynchronous secondaries are absent among observed binary systems with close orbits ($a/D_1 \lesssim 2.2$, $P_{\rm orb} \lesssim 20$ h). They also pointed out that asynchronous secondaries are typically observed on eccentric orbits. Based on that, the parameters of the mutual orbit of 2001 SL9 (a close orbit with low or zero eccentricity) and the fact that the secondary spin relaxation is typically faster than the orbit circularization (Goldreich and Sari, 2009), we assume that the secondary of 2001 SL9 is in synchronous rotation, i.e., its rotation period is the same as the orbit period.

Earlier work where some of the binary asteroid parameters were derived is Pravec et al. (2006). Their values ($P_1 = 2.4004 \pm 0.0002$ h, $P_{\rm orb} = 16.40 \pm 0.02$ h, $D_{2,\rm C}/D_{1,\rm C} = 0.28 \pm 0.02$, $D_{1,\rm C} = 0.8$ km uncertain to a factor of two) are generally in agreement with our current best estimated parameters, but they did not perform a modeling in order to get parameters of the mutual orbit.

(88710) 2001 SL9 appears to be a typical near-Earth binary asteroid according to its basic parameters. Its bulk density of $\sim 1.8$ g cm$^{-3}$ is in good agreement with its rocky taxonomical class. The normalized total angular momentum content of 2001 SL9 is $\alpha_L = 1.1 \pm 0.2$ (1-$\sigma$ uncertainty), i.e., in the range 0.9–1.3 for small near-Earth and main belt asteroid binaries and exactly as expected for the

Table 5
Properties of binary asteroid (88710) 2001 SL9.

| Parameter | | Value | Unc. | Reference |
|---|---|---|---|---|
| Whole system: | | | | |
| $H_V$ | | $17.98 \pm 0.02$ | $1\sigma$ | This work |
| $G$ | | $0.34 \pm 0.03$ | $1\sigma$ | This work |
| $V - R$ | | $0.457 \pm 0.010$ | $1\sigma$ | This work |
| $D_{\rm eff}$ (km) | | $0.75 \pm 0.10^a$ | $1\sigma$ | This work |
| Taxon. class | | Sr, Q | | P18, L05 |
| Primary: | | | | |
| $D_{1,\rm C}$ (km) | | $0.73 \pm 0.32$ | $3\sigma$ | This work |
| $D_{1,\rm V}$ (km) | | $0.77 \pm 0.34$ | $3\sigma$ | This work |
| $P_1$ (h) | | $2.4004 \pm 0.0002$ | $1\sigma$ | P06 |
| $(A_1B_1)^{1/2}/C_1$ | | $\leq 2.2^b$ | $3\sigma$ | This work |
| $A_1/B_1$ | | $1.07 \pm 0.01$ | $1\sigma$ | PH07 |
| $\rho_1 = \rho_2$ (g cm$^{-3}$) | | $1.8^{+2.5}_{-0.5}$ | $3\sigma$ | This work |
| Secondary: | | | | |
| $D_{2,\rm C}/D_{1,\rm C}$ | | $0.24 \pm 0.02$ | $3\sigma$ | This work |
| $D_{2,\rm C}$ (km) | | $0.18 \pm 0.08$ | $3\sigma$ | This work |
| $D_{2,\rm V}$ (km) | | $(0.18 \pm 0.08)^c$ | $3\sigma$ | This work |
| $P_2$ (h) | | $(16.40)^d$ | | |
| $A_2/B_2$ | | $\leq 1.2$ | $3\sigma$ | This work |
| Mutual orbit: | Solution | | | |
| $a/(A_1B_1)^{1/2}$ | | $1.75 \pm 0.3$ | $3\sigma$ | This work |
| $(L_{\rm P}, B_{\rm P})$ (deg.) | | $(302, -73) \pm (10 \times 4)^e$ | $3\sigma$ | This work |
| $P_{\rm orb}$ (h) | 1. | $16.4022 \pm 0.0002^f$ | $3\sigma$ | This work |
| | 2. | $16.4027 \pm 0.0002^f$ | | |
| $L_0$ (deg.) | 1. | $51 \pm 5^f$ | $3\sigma$ | This work |
| | 2. | $56 \pm 5^f$ | | |
| $e$ | | $\leq 0.07$ | $3\sigma$ | This work |
| $\Delta M_d$ (deg/yr$^2$) | 1. | $2.8 \pm 0.2$ | $3\sigma$ | This work |
| | 2. | $5.2 \pm 0.2$ | | |
| $\dot{P}_{\rm orb}$ (h/yr) | 1. | $-0.00048 \pm 0.00003$ | $3\sigma$ | This work |
| | 2. | $-0.00089 \pm 0.00004$ | | |
| $\dot{a}$ (cm/yr) | 1. | $-2.8 \pm 0.2$ | $3\sigma$ | This work |
| | 2. | $-5.1 \pm 0.2$ | | |

References: L05 (Lazzarin et al., 2005), P18 (Pajuelo et al., 2018), P06 (Pravec et al., 2006), PH07 (Pravec and Harris, 2007).
[a] From the derived $H_V$ and assumed $p_V = 0.20 \pm 0.05$ that is the mean albedo for S-complex asteroids (Pravec et al., 2012).
[b] The formal best-fit value is 1.7.
[c] Assuming a spherical shape of the secondary.
[d] The secondary is assumed to in synchronous rotation. See text for details.
[e] These are the semiaxes of the uncertainty area; see its actual shape in Fig. 12.
[f] These are the periods and $L_0$ for epoch JD 2456182.39026 (asterocentric time, i.e., light-time corrected).

proposed formation of small binary asteroids by fission of critically spinning rubble-pile progenitors (Pravec and Harris, 2007).

According to JPL HORIZONS system, the closest Earth, Venus and Mars approaches of 2001 SL9 from 2001 to 2015 were 0.22, 0.13 and 0.36 AU, respectively. We can therefore rule out planetary-tug effects as a potential cause for the observed mutual orbital period decrease.

## 5 Implications for the BYORP effect

### *5.1 (66391) 1999 KW4 BYORP Modeling*

McMahon and Scheeres (2010b) computed a BYORP coefficient, $B$, for the secondary shape of 1999 KW4 based on the model published by Ostro et al. (2006). The nominal coefficient was found to be $B_{nom} = 2.082\times10^{-2}$, and based on the other parameters of the system this produced a semimajor axis drift rate of approximately 7 cm/yr, according to the relationship

$$\dot{a}_B = \frac{2P_\Phi}{a_h^2\sqrt{1-e_h^2}} \frac{a^{3/2}R_{mean}^2}{m_s\sqrt{\mu}} B, \tag{3}$$

where $P_\Phi$ is the solar radiation pressure constant, whose value is taken to be $10^{14}$ kg km/s$^2$; $a_h$ is the heliocentric orbit semimajor axis of 0.642 AU, and $e_h$ = 0.688 is the heliocentric orbit eccentricity. The other values – binary orbit semimajor axis, $a$, secondary mass, $m_s$, binary gravitational parameter, $\mu$ can be obtained from Table 4. The secondary mean radius, $R_{mean}$ was computed as average of vertices of the shape model of the secondary from Ostro et al., scaled up by 130%. The secondary mass can be expressed in terms of the estimated secondary volume, $V_s$ and bulk density, $\rho_s$. The values are: $V_s = 0.105$ km$^3$, $\mu = 131.5$ m$^3$/s$^2$, $a = 2.548$ km, $R_{mean} = 283.7$ m. [3]

For $\rho_s$ we used a value of 1.97 g/cm$^3$ – the density of the primary from Ostro et al., assuming that both components have the same density. The BYORP modeling with these newly estimated parameters gives $\dot{a} = 8.53$ cm/year, which is significantly larger than the observed value of 1.2 cm/year.

Given the previous discussion of the uncertainty in the secondary shape from Ostro et al. (2006), and the fact that we find an increase in size of approximately 30%, it is reasonable to assume that many details of the shape may not be accurately known. If the topography changes, the predicted BYORP coefficient will also change. To investigate this, we modeled the predicted BYORP effect for a suite of shapes similar to the KW4 secondary radar shape model from Ostro et al. (2006), to compute the likely range of values for the BYORP effect, using the computational model of McMahon and Scheeres (2010a), which incorporates self-shadowing and secondary intersections of re-radiated energy. The shapes were changed by perturbing the vertices radially [4] using the random Gaussian spheroid method (Muinonen 2010). The radial perturbations were set to approximate the size estimate accuracy given in Ostro et al. (2006) of 6% of the long axis, which comes out to 17.1 m for a 1$\sigma$ radial dispersion. The correlation distances were set as 50 m (making small scale, "spiky" topography features), 150 m, and 300 m (smoother global variations in topography).

[3] Note that the mean radius ($R_{mean}$) used to normalize the BYORP coefficient ($B$) is the mean of the vertex radii of the shape model. Upon scaling the KW4 secondary shape model up by 130% here, the mean radius to be used in Eq. (3) is 283.7 m.

[4] The vector from the origin to the vertex keeps its direction, but changes its magnitude.

The BYORP coefficients were computed for 90 such randomly perturbed shapes for each correlation distance.

The results of this process can be seen in Fig. 14. As can be seen, these drift rates are all still higher than the measured value. The associated BYORP coefficients range from $B_{min} = 7.701 \times 10^{-3}$ to $B_{max} = 3.323 \times 10^{-2}$.

One other parameter that is poorly constrained in Eq. 3 is the secondary density. In fact, while the previous computations assumed an equal density across both components, Ostro et al (2006) reported its large uncertainty. The effect of a variation in secondary density (with total system mass $\mu$ being held constant) can be seen in Fig. 15. It can be seen here that in order for the secondary density alone to modify the semimajor axis drift rate to match the measured value – even with the minimum BYORP coefficient seen – the secondary density would have to be approximately 3.6 g/cm$^3$ – significantly higher than Ostro's estimate, and requiring a significantly more dense secondary than primary, but not impossible in terms of bulk density alone.

The total semimajor axis drift rate for a binary asteroid is governed by the interplay between BYORP and tides, however tides are always expansive for 1999 KW4. The tide induced semimajor axis drift rates can be computed (Jacobson and Scheeres, 2011) as

$$\dot{a}_T = 3\frac{k_p}{Q}\left(\frac{\omega_d}{a_{rp}^{11/2}}\right) q\sqrt{1+q}, \tag{4}$$

where the surface disruption spin limit for a sphere is given by

$$\omega_d = (4\pi G\rho/3)^{1/2} \tag{5}$$

and $q$ is the mass ratio, $k_p$ is the tidal Love number of the primary, $Q$ is the tidal dissipation number, and $G$ is the gravitational constant, and $a_{rp} = a/(D_{1,\mathrm{V}}/2)$ is the binary semimajor axis in units for primary radii. For the current estimate of 1999 KW4, $q = m_s/m_p = 0.090$, $\omega_d = 6.501 \times 10^{-4}$ rad/s, and the primary radius is taken to be $R_p = 0.6585$ km.

$Q/k_p$ is a relatively unknown parameter for rubble pile asteroids, but two values have emerged from the literature: $2.15 \times 10^7$ (Taylor and Margot, 2011; recomputed from their value of $\mu Q$ derived for 1999 KW4) and $2.40 \times 10^5$ (Scheirich et al. 2015; derived for 1996 FG3). Since the theory and observations suggest that $Q/k$ scale with $R_p$ (Nimmo and Matsuyama, 2019), we scaled the second value (from 1996 FG3) using diameter of 1999 KW4 to $1.87 \times 10^5$. We note that Taylor and Margot (2011) assumed a maximum tidal evolution timescale and so their $Q/k$ is a lower bound.

Using these two values as bounds, we find that the tide induced semimajor axis drift rate ranges from $\leq 0.0158$ to 1.813 cm/yr.

The semimajor axis drift rate from BYORP is also lowered if the secondary is librating significantly, however the lightcurve observations show little evidence of this, implying that if there is any libration it is small and the degradation in the drift rate would be minimal. Thus, the overall BYORP coefficient may be significantly lower than predicted from our direct geometric theory or have an opposite sign, implying that the system may be moving into an equilibrium.

Given the observed semimajor axis drift rate (1.2 cm/year), we also estimate the BYORP coefficient for a case if the tides were insignificant and for the two values of $Q/k_p$ mentioned above. To explain the observed drift rate due to the BYORP alone, the BYORP coefficient would need to be $B = 0.003$. Considering the tides as well, using $Q/k_p \geq 2.15 \times 10^7$ we obtain $B \geq 0.003$. Using $Q/k_p = 1.87 \times 10^5$ we obtain $B = -1.57 \times 10^{-3}$.

### *5.2 (88710) 2001 SL9 BYORP Modeling*

Unlike with 1999 KW4, there is no detailed shape model available for the secondary of 2001 SL9 (except for an upper limit on its elongation, which is insufficient for BYORP calculation), so that no informed forward modeling for the BYORP coefficient can be carried out. Instead, we compute the value of the BYORP coefficient that would produce the measured semimajor axis drift rates.

Given that the secondary is assumed to be in synchronous rotation while the primary is spinning much faster than the orbit period, the tides work to expand the semimajor axis. Thus, inward BYORP must overcome tides to achieve the measured semimajor axis rates. Due to the uncertainty in the $Q/k$, we report four possible BYORP coefficients for 2001 SL9 in Table 6 – one for each combination of drift rate and tidal parameters. The first value of $Q/k = 1.1 \times 10^5$ was scaled from the value derived by Scheirich et al. (2015) for 1996 FG3, the second value of $Q/k \geq 2.5 \times 10^7$ was recomputed from $\mu Q$ derived by Taylor and Margot (2011) for 2001 SL9.

Table 6
Computed BYORP coefficient, $B$, for SL9 based on measured semimajor axis drift rates and possible $Q/k$ values.

| | $B$ | |
|---|---|---|
| | $Q/k = 1.1 \times 10^5$ | $Q/k \geq 2.5 \times 10^7$ |
| $\dot{a}$ = -2.8 cm/yr | $-5.92 \times 10^{-3}$ | $\leq -6.18 \times 10^{-3}$ |
| $\dot{a}$ = -5.1 cm/yr | $-1.10 \times 10^{-2}$ | $\leq -1.12 \times 10^{-2}$ |

Note that for 2001 SL9, $q = 0.0128$, $\omega_d = 7.094 \times 10^{-4}$ rad/s, and given the primary radius of 0.385 km ($= D_{1,V}/2$), we get $a_{rp} = 4.177$. It is important to point out that BYORP is the only known physical mechanism that can cause an inward semimajor axis drift rate, as measured here for 2001 SL9. The computed $B$ coefficient magnitudes are in line with the modeled values for 1999 KW4, providing some confidence that the results are reasonable.

The results shown here, combined with the BYORP-tide equilibrium state detected for 1996 FG3[5] (Scheirich et al., 2015) does imply that BYORP effect seems to be real, but that we cannot adequately compute it as of yet. This inadequacy could either be from error in the shape models or a deficit in the theory.

### *5.3 Differential Yarkovsky force in binary asteroid system*

Another effect affecting the magnitude of the mutual semimajor axis drift is the Yarkovsky force, which affects not only the motion of the center of mass of the whole binary system but also the relative motion of components. We computed the effect by a method described in Vokrouhlický et al. (2005). The shapes of the components were approximated by spheres represented by regular polyhedrons with 504 surface elements. The Yarkovsky accelerations $\mathbf{f_1}$ and $\mathbf{f_2}$ of both components were determined by numerical solution of the heat diffusion problem. The accelerations for the two components differ because of different sizes and spin rates. Moreover, they are affected by mutual shadowing of the components. Assuming zero eccentricity, the drift of the semimajor axis of the mutual orbit is $\dot{a} = 2/n \, \langle f_\tau \rangle$, where $n$ represents the mean motion and $\langle f_\tau \rangle$ is a heliocentric-orbit averaged value of $f_\tau$ – a projection of the difference between the two Yarkovsky accelerations to the transversal direction of the relative motion $\mathbf{e}_\tau$,

$$f_\tau = \mathbf{e}_\tau \cdot (\mathbf{f_2} - \mathbf{f_1}). \tag{6}$$

Without the mutual shadowing of the components the value of $\langle f_\tau \rangle$ would be zero. Therefore, the resulting mutual semimajor axis drift depends also on the orientation of the heliocentric and mutual orbits.

The Yarkovsky acceleration is less sensitive to body's shape than to its thermophysical parameters. The results for the semimajor axis drift are shown in Fig. 16. For the nominal solution of (66391) 1999 KW4 and the thermal inertia range 100–1000 $\mathrm{J\,m^{-2}\,s^{-1/2}\,K^{-1}}$ (Delbo et al., 2015), the semimajor axis drift is between $-4$ mm and $-8$ mm per year. With the pole of mutual orbit inside the admissible area (see Fig. 7), the drift of mutual semimajor axis can differ by a factor of $\sim$2 from the value for nominal solution.

In the case of (88710) 2001 SL9, the Yarkovsky force has only negligible effect on the mutual semimajor axis drift. For the nominal parameters the drift is $\sim -1$ mm/yr (see Fig. 16). Depending on the orientation of the mutual orbit within its admissible area, the value can differ by a factor of $\sim$2.

[5] In Scheirich et al., 2015, we did not include the length of the secondary, $L_0$, in Table 3. We complete the information here: its value is $56 \pm 6$ deg for epoch JD 2450183.47442 (asterocentric time, i.e., light-time corrected).

## 6 Conclusions

The near-Earth asteroids (66391) 1999 KW4 and (88710) 2001 SL9 are among the best characterized small binary asteroid systems. They are typical members of the population of near-Earth asteroid binaries for most of their parameters. With the data from our photometric observations taken during six apparitions over the time interval of 19 years and during five apparitions over almost 14 years for (66391) and (88710), respectively, we constrained the long-term evolution of their binary orbits.

For (66391), we found that the semimajor axis of its mutual orbit is expanding with a rate of $1.2 \pm 0.3$ cm/yr ($3\sigma$). The observed drift is on an order of the theoretical drift rate caused by mutual tides ($\leq 0.0158$ to 1.813 cm/yr). However the predicted drift caused by the BYORP effect (8.53 cm/yr) is much higher than the observed value. Thus, the BYORP coefficient may be significantly lower than predicted from a direct geometric theory by McMahon and Scheeres (2010a) or have an opposite sign, implying that the system may be moving into an equilibrium.

For (88710), we found that the semimajor axis of its mutual orbit is shrinking with a rate of $-2.8 \pm 0.2$ or $-5.1 \pm 0.2$ cm/yr ($3\sigma$). The BYORP effect is the only known physical mechanism (except the differential Yarkovsky effect, which is much slower than the observed value) that can cause an inward drift. Since there is no shape model available for the secondary, no forward modeling for the BYORP coefficient is possible. Instead, the BYORP coefficient can be computed from the measured drift rates. The computed coefficient magnitudes are similar to the modeled values for (66391) 1999 KW4, providing some confidence that the results are reasonable.

### Acknowledgements

The work at Ondřejov Observatory and observations with the Danish 1.54-m telescope on the ESO La Silla station were supported by the Grant Agency of the Czech Republic, Grant 20-04431S, and by the project RVO:67985815. Access to computing and storage facilities owned by parties and projects contributing to the National Grid Infrastructure MetaCentrum provided under the program "Projects of Large Research, Development, and Innovations Infrastructures" (CESNET LM2015042), and the CERIT Scientific Cloud LM2015085, is greatly appreciated. The observations at Maidanak Observatory were supported by grant VA-FA-F-2-010 of the Ministry of Innovative Development of Uzbekistan. The work at Abastumani was supported by the Shota Rustaveli National Science Foundation, Grant RF-18-1193. The authors acknowledge the sacred nature of Mauna Kea, and appreciate the opportunity to observe from the mountain. The authors would like to thank the University of Hawaii for using the 2.2 m telescope.

### Appendix A. Taxonomic classification of (66391) 1999 KW4

Near-infrared (NIR) spectra (0.7-2.5 $\mu$m) of (66391) 1999 KW4 were obtained in low-resolution prism mode on May 28, 2019 (5:30–7:55 UTC) with the SpeX instrument (Rayner et al., 2003) on NASA Infrared Telescope Facility (IRTF). The asteroid was 12.6 visual magnitude and was observed at a phase angle of 81°, and an airmass of ∼1.2-1.7. Weather conditions were stable during the observing run, with a seeing of 0.9" and a humidity of ∼25%. During the observations, the 0.8"-slit was oriented along the parallactic angle in order to minimize the effects of differential atmospheric refraction. To avoid saturation, the integration time was limited to 60 seconds. A G-type local extinction star was observed before and after the asteroid in order to correct the telluric bands. Solar analog SAO 120107 was also observed to correct for possible spectral slope variations. All spectra were reduced using the IDL-based software Spextool (Cushing et al., 2004). A detailed description of the steps involved in the data reduction process can be found in Sanchez et al. (2013).

The average NIR spectrum of (66391) 1999 KW4 is shown in Fig. 17. The spectrum exhibits two very deep absorption bands at 0.94 and 1.94 $\mu$m, due to the presence of olivine and pyroxene. Using the online Bus-DeMeo taxonomy calculator (http://smass.mit.edu/busdemeoclass.html) we found that (66391) is classified as either O- or Q-type in this taxonomic system (DeMeo et al., 2009). A visual inspection shows that the overall spectral characteristics of (66391) are more similar to a Q-type asteroid. However, we noticed that the absorption bands in the NIR spectrum of (66391) are much deeper than those of a typical Q-type. Band depths are measured from the continuum to the band centers and are given as percentage depths (Clark and Roush, 1984). For (66391), we found that the Band I depth is $34.4 \pm 0.2\%$, and the Band II depth is $15.9 \pm 0.2\%$, while the mean spectrum of a Q-type asteroid (DeMeo et al. 2009) has Band I and II depths of $23.8 \pm 0.1\%$, and $6.0 \pm 0.2\%$, respectively. This difference could be attributed to several factors, including mineral abundance, grain size, and the high phase angle at which (66391) was observed (e.g., Sanchez et al. 2012).

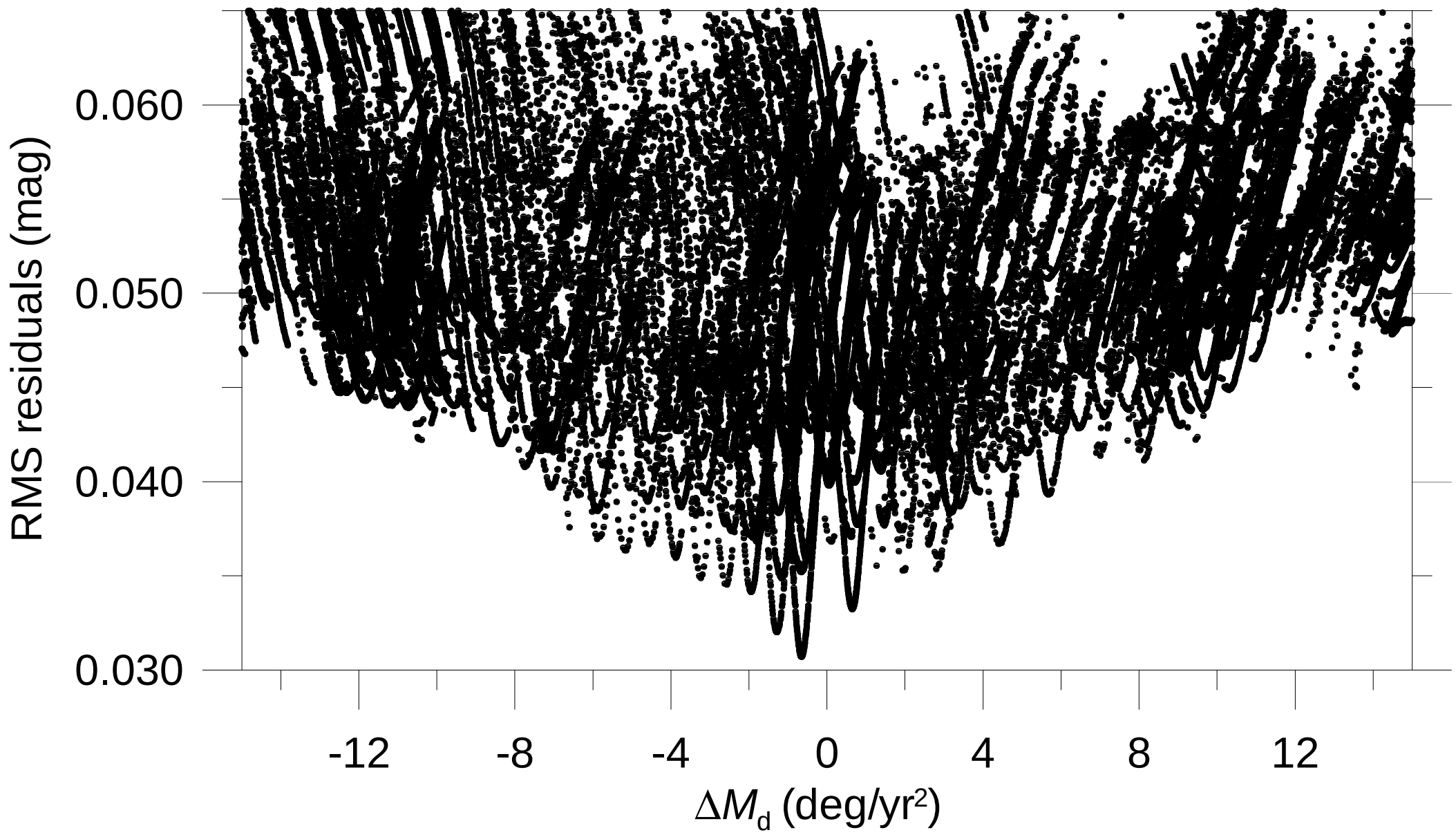

Fig. 1. The RMS residuals vs. $\Delta M_d$ for solutions of the model of (66391) 1999 KW4 presented in Section 2.2. Each dot represents the best-fit result with $\Delta M_d$ fixed and other parameters varied. The plots were constructed using spherical shapes of both components; see text for details.

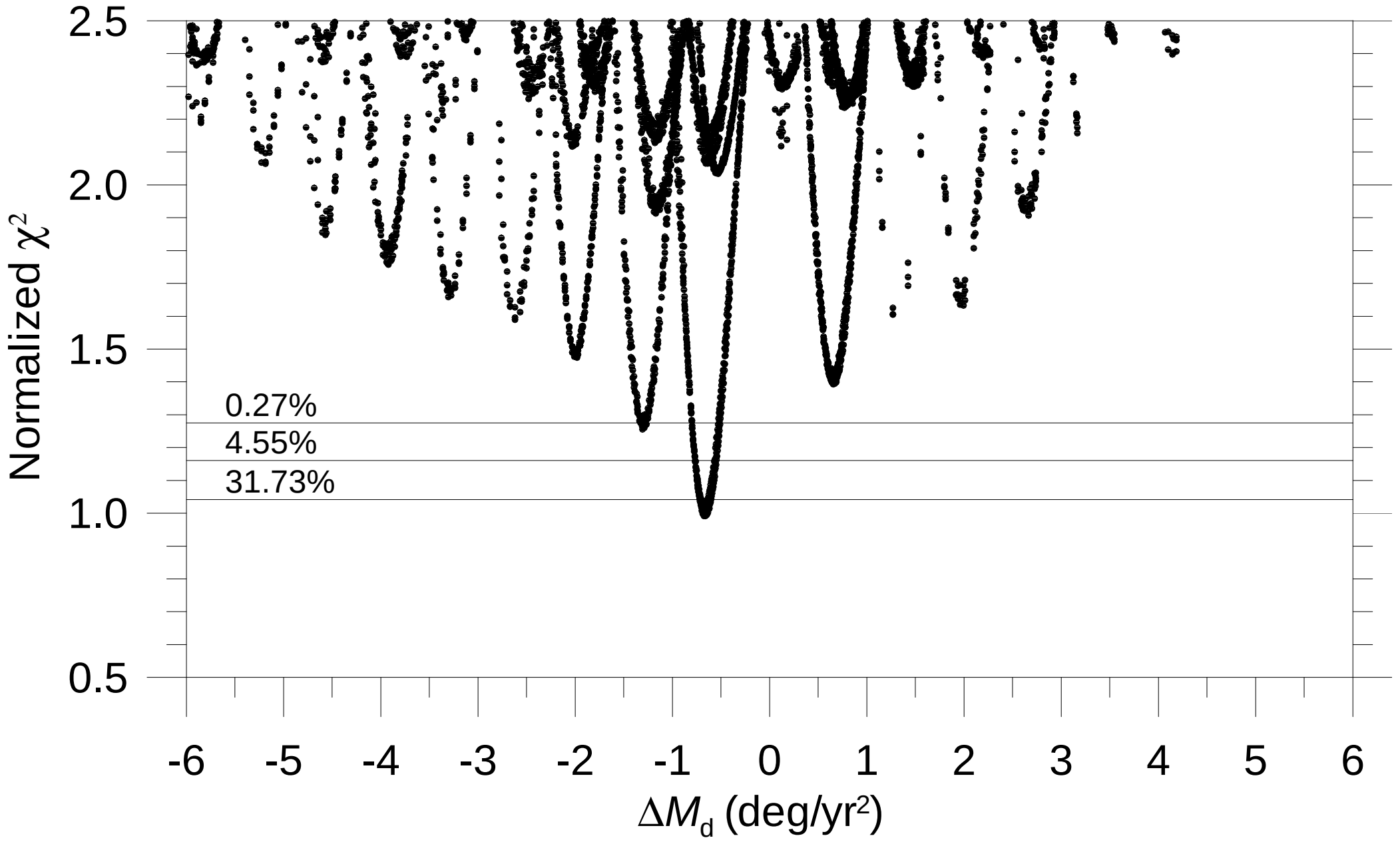

Fig. 2. The normalized $\chi^2$ vs. $\Delta M_d$ for solutions of the model of (66391) 1999 KW4 presented in Section 2.2. The three horizontal lines gives the p-values – the probabilities that the $\chi^2$ exceeds a particular value only by chance, corresponding to 1-, 2- and 3$\sigma$ interval of the $\chi^2$ distribution with 236 degrees of freedom. See text for details.

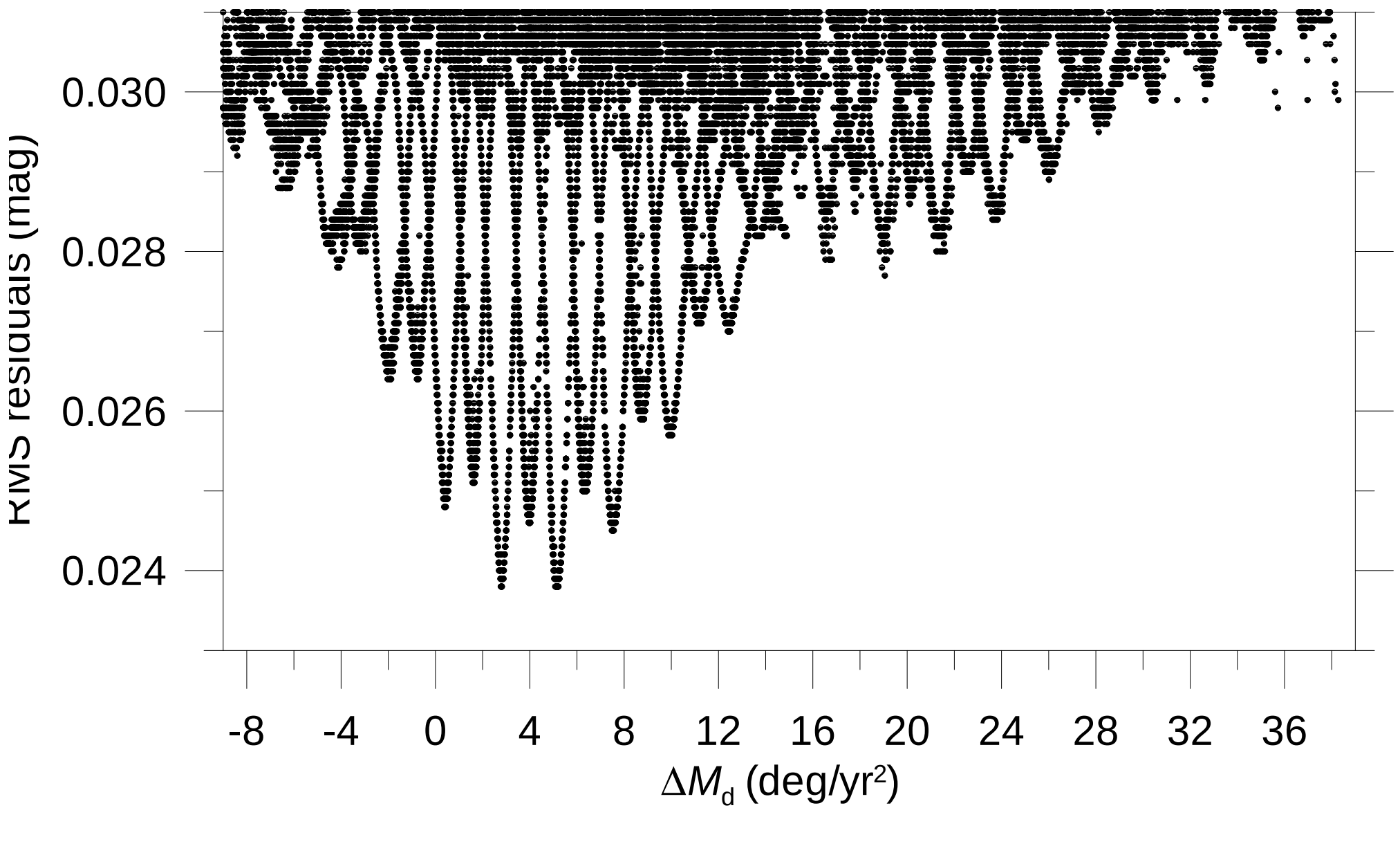

Fig. 3. The RMS residuals vs. $\Delta M_d$ for solutions of the model of (88710) 2001 SL9 presented in Section 2.2. Each dot represents the best-fit result with $\Delta M_d$ fixed and other parameters varied. The plots were constructed using spherical shapes of both components; see text for details.

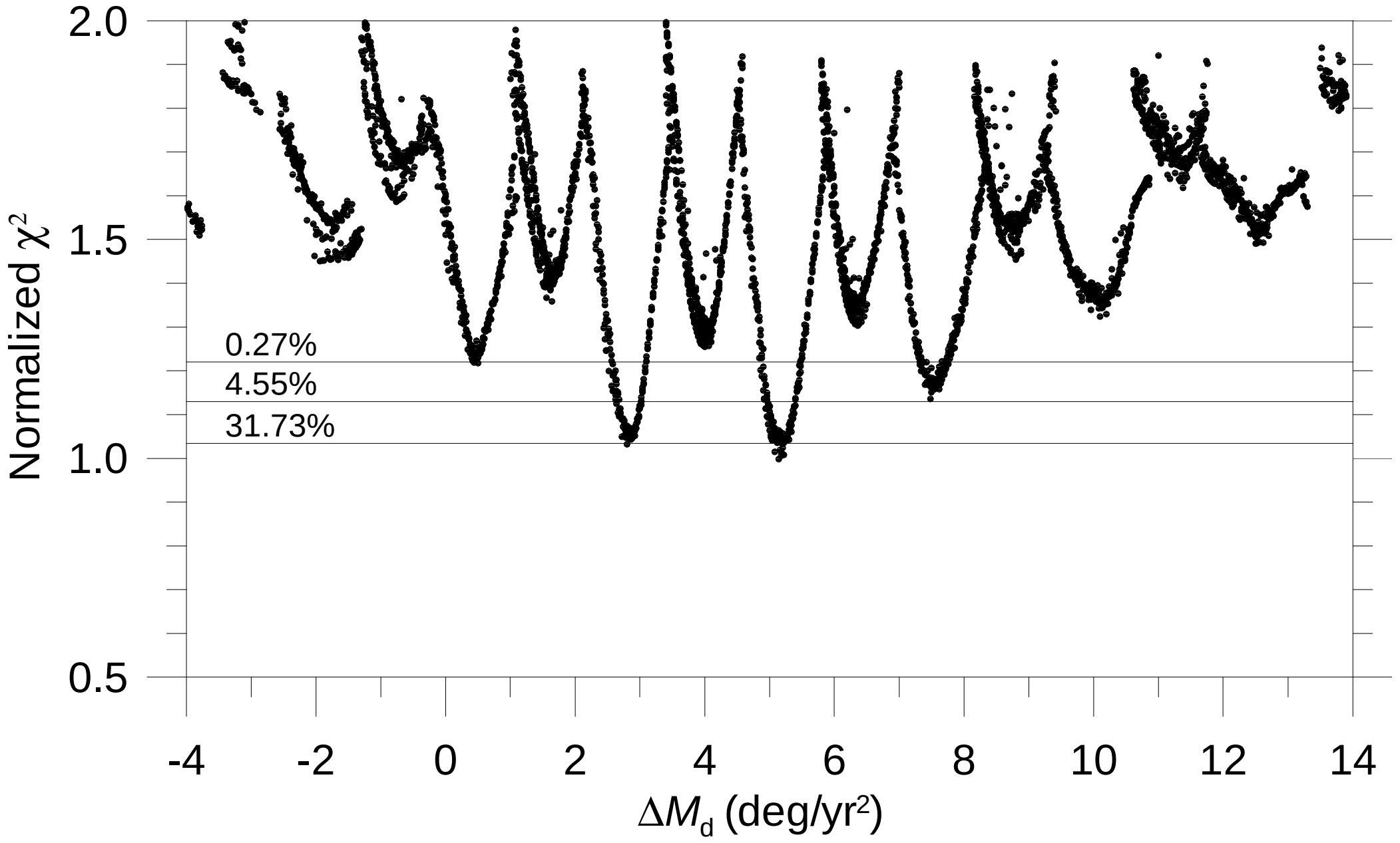

Fig. 4. The normalized $\chi^2$ vs. $\Delta M_d$ for solutions of the model of (88710) 2001 SL9 presented in Section 2.2. The three horizontal lines gives the p-values – the probabilities that the $\chi^2$ exceeds a particular value only by chance, corresponding to 1-, 2- and 3$\sigma$ interval of the $\chi^2$ distribution with 359 degrees of freedom. See text for details.

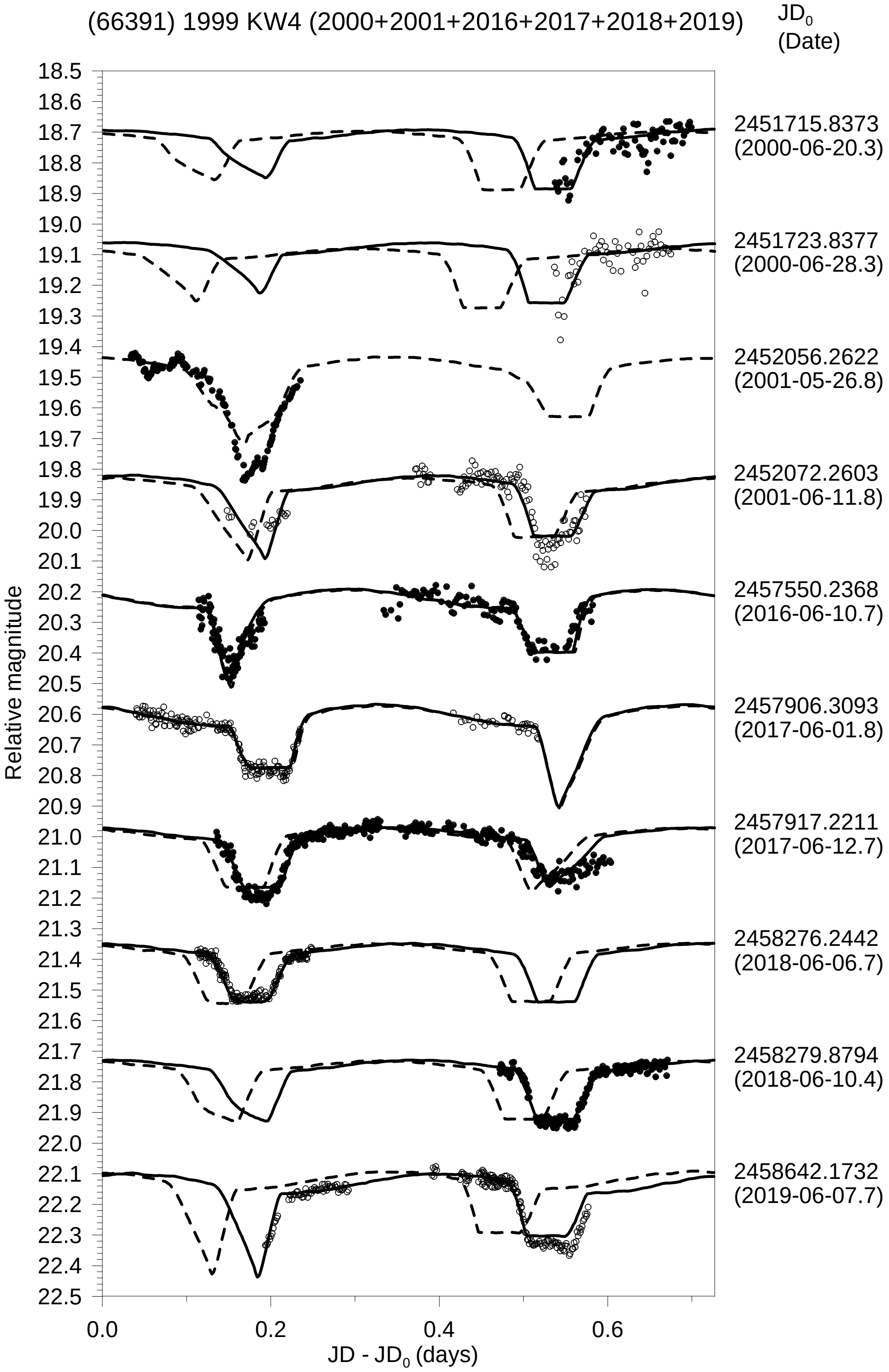

Fig. 5. Selected data of the long-period lightcurve component of (66391) 1999 KW4. The observed data are marked as points. The solid curve represents the synthetic lightcurve for the best-fit solution with $\Delta M_d = -0.65$ deg/yr$^2$. For comparison, the dashed curve is the model with $\Delta M_d$ fixed at 0.0 deg/yr$^2$ and all other parameters varied to obtain the best fit.

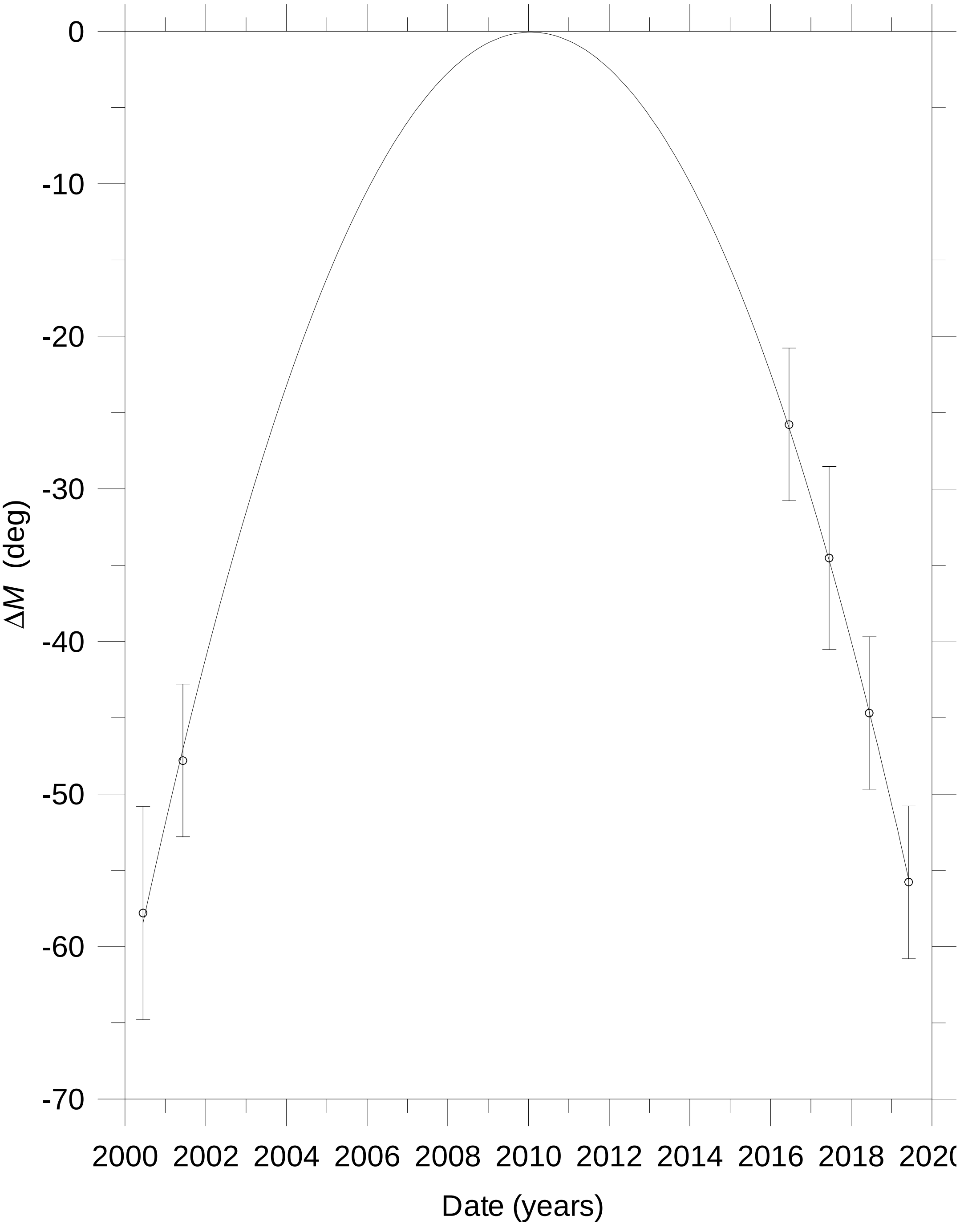

Fig. 6. A time evolution of the mean anomaly difference $\Delta M$ for (66391) 1999 KW4. See text for details. Each point corresponds to the middle of one of the six apparitions from 2000 to 2019. Vertical error bars represent estimated $3\sigma$ uncertainties of the event times, expressed in mean anomaly. The solid curve is a quadratic fit to the data points.

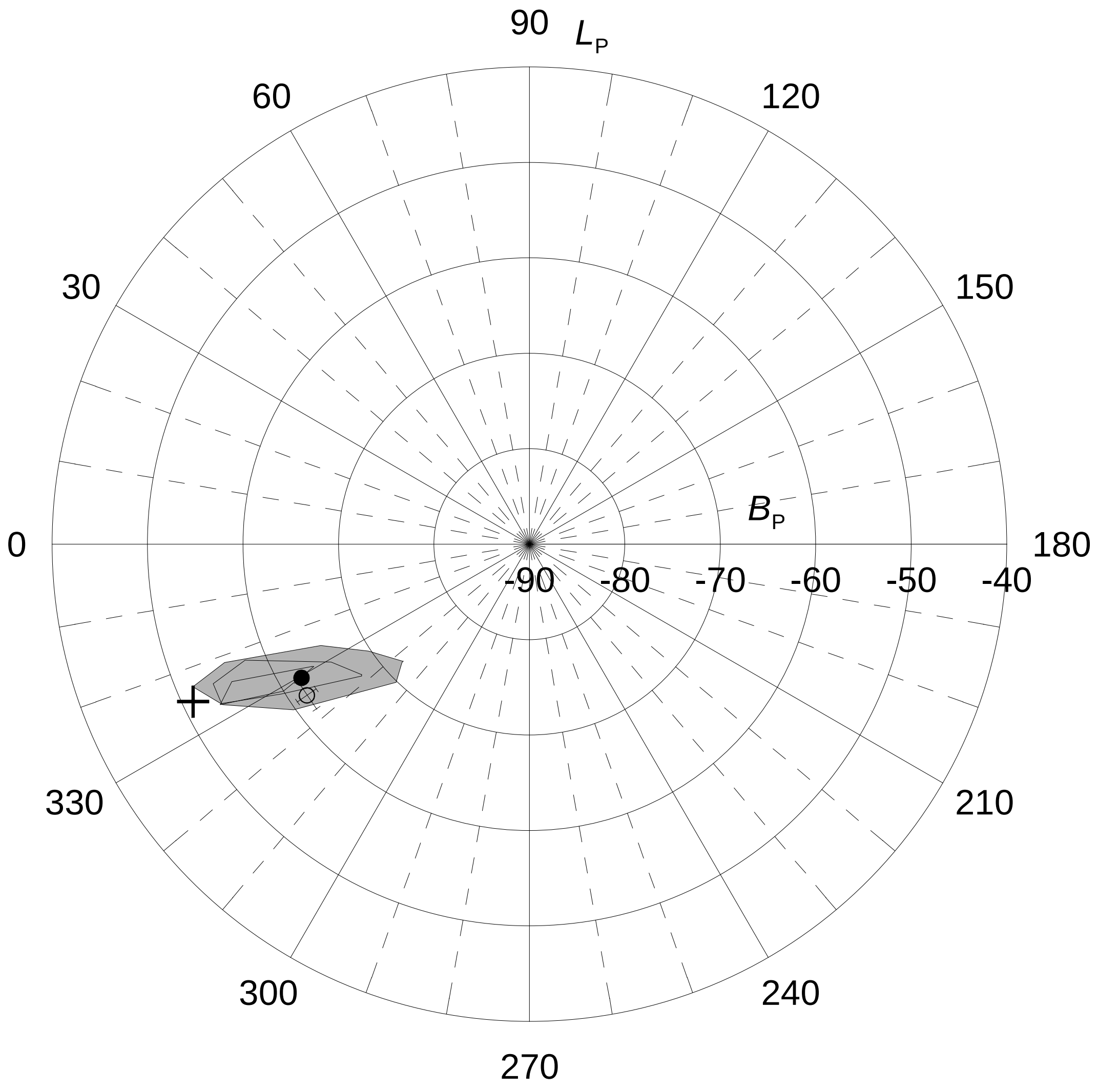

Fig. 7. Area of admissible poles for the mutual orbit of (66391) 1999 KW4 in ecliptic coordinates (grey area) for $(A_1B_1)^{1/2}/C_1 = 1$. The dot is the nominal solution given in Table 4. This area corresponds to $3\sigma$ confidence level. To demonstrate the effect of a flattening of the primary on the estimated pole, the areas confined by solid lines shows the admissible poles constrained using $(A_1B_1)^{1/2}/C_1 = 1.2$ (middle area) and 1.4 (the smallest area). The open circle with error bars represents a solution for the orbital pole from Ostro et al. (2006) with $1\sigma$ uncertainties. The south pole of the current asteroid's heliocentric orbit is marked with the cross.

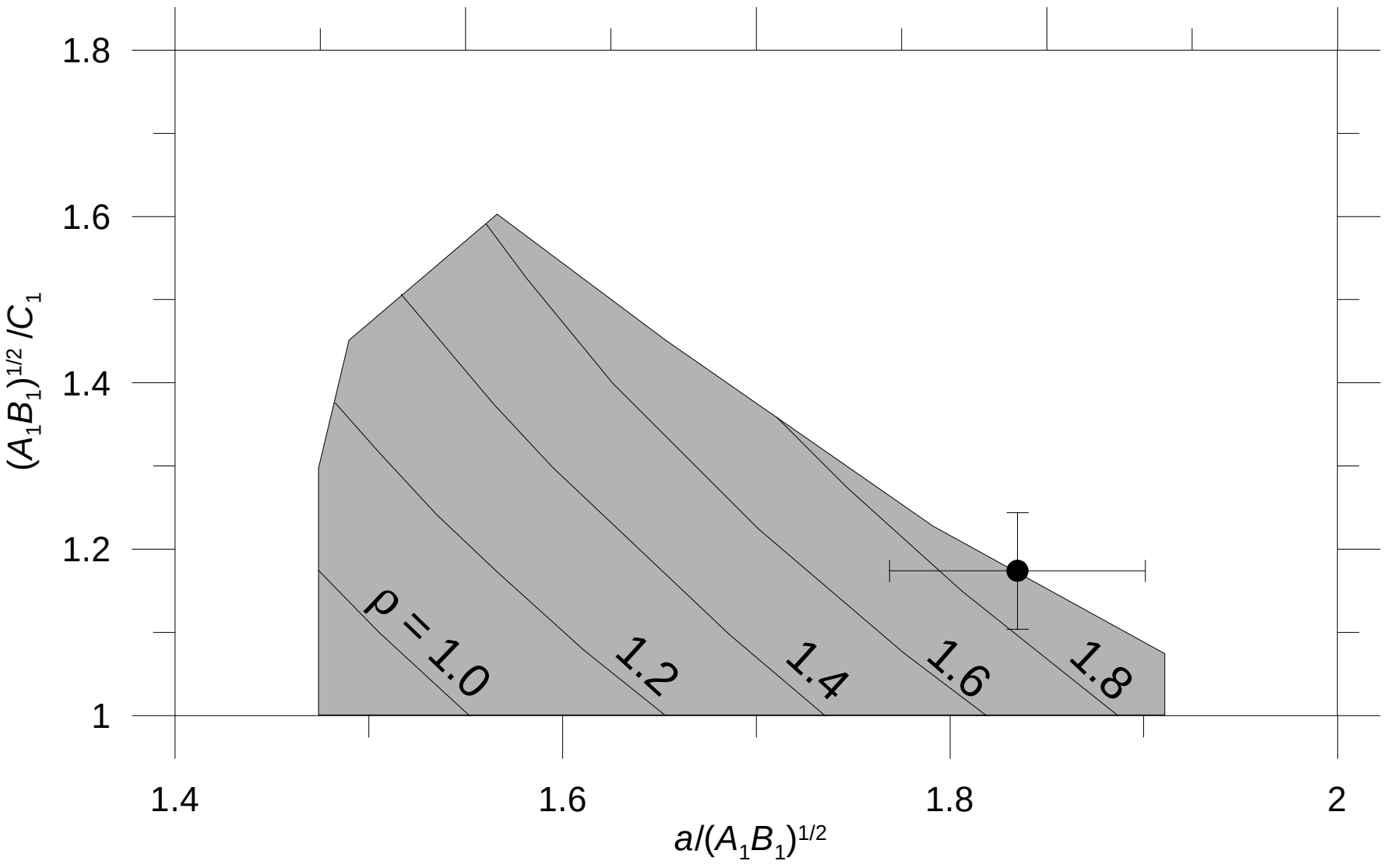

Fig. 8. Area of admissible combinations of the ratio between the mean equatorial and the polar axes of the primary ($(A_1B_1)^{1/2}/C_1$) and the semimajor axis of the mutual orbit $a$ of (66391) 1999 KW4. This area corresponds to $3\sigma$ confidence level. Values of the bulk density of the system ($\rho$) in g cm$^{-3}$ are indicated. The dot with the error bars is the result from Ostro et al. (2006) and its $1\sigma$ uncertainties (see text for details).

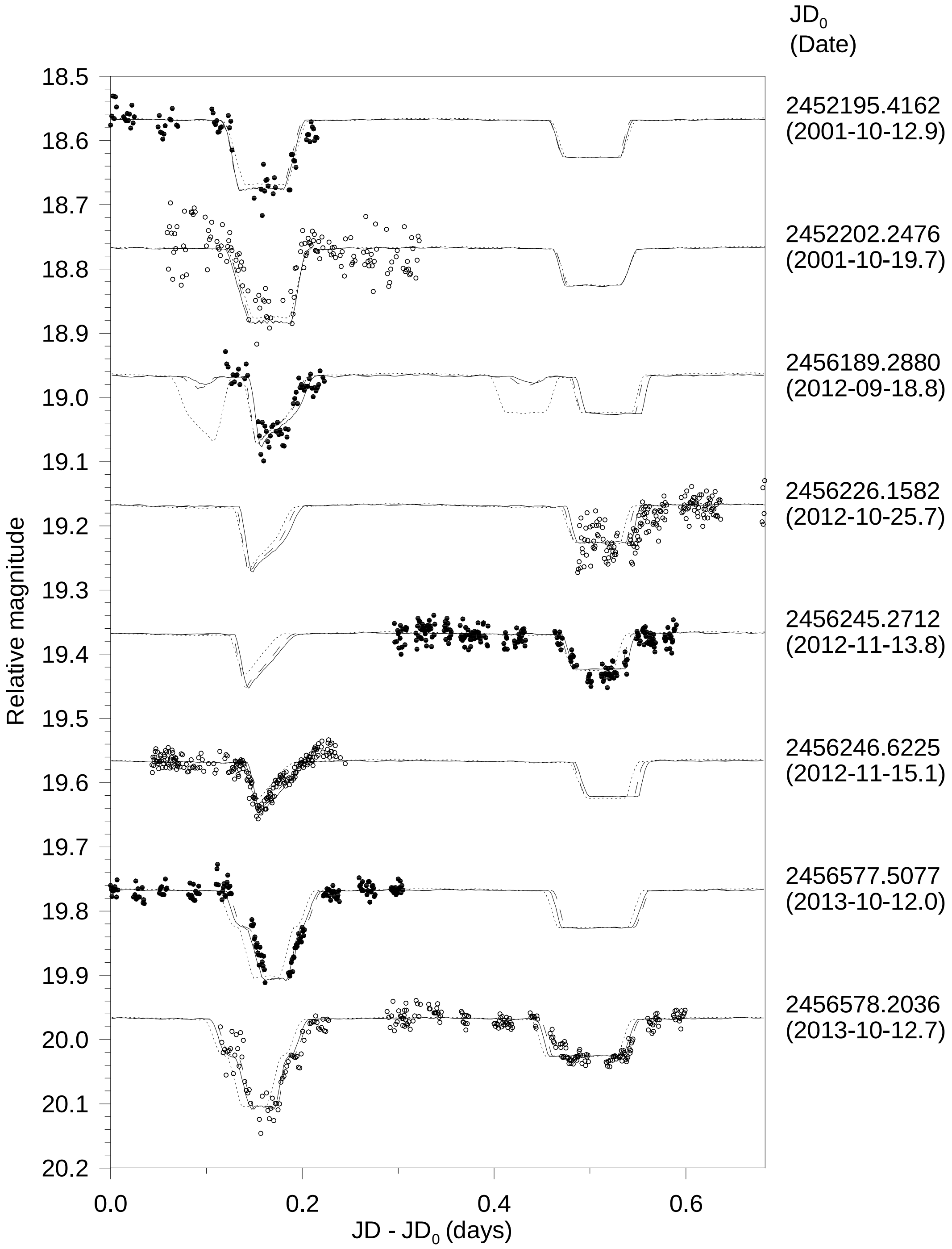

Fig. 9. Selected data of the long-period lightcurve component of 2001 SL9. The observed data are marked as points. The solid and dashed curves represent the synthetic lightcurves of the two best-fit solutions with $\Delta M_d = 2.8$ and $5.2$ deg/yr$^2$, respectively. For comparison, the dotted curve is for the best-fit model with $\Delta M_d$ fixed at 0.0 deg/yr$^2$.

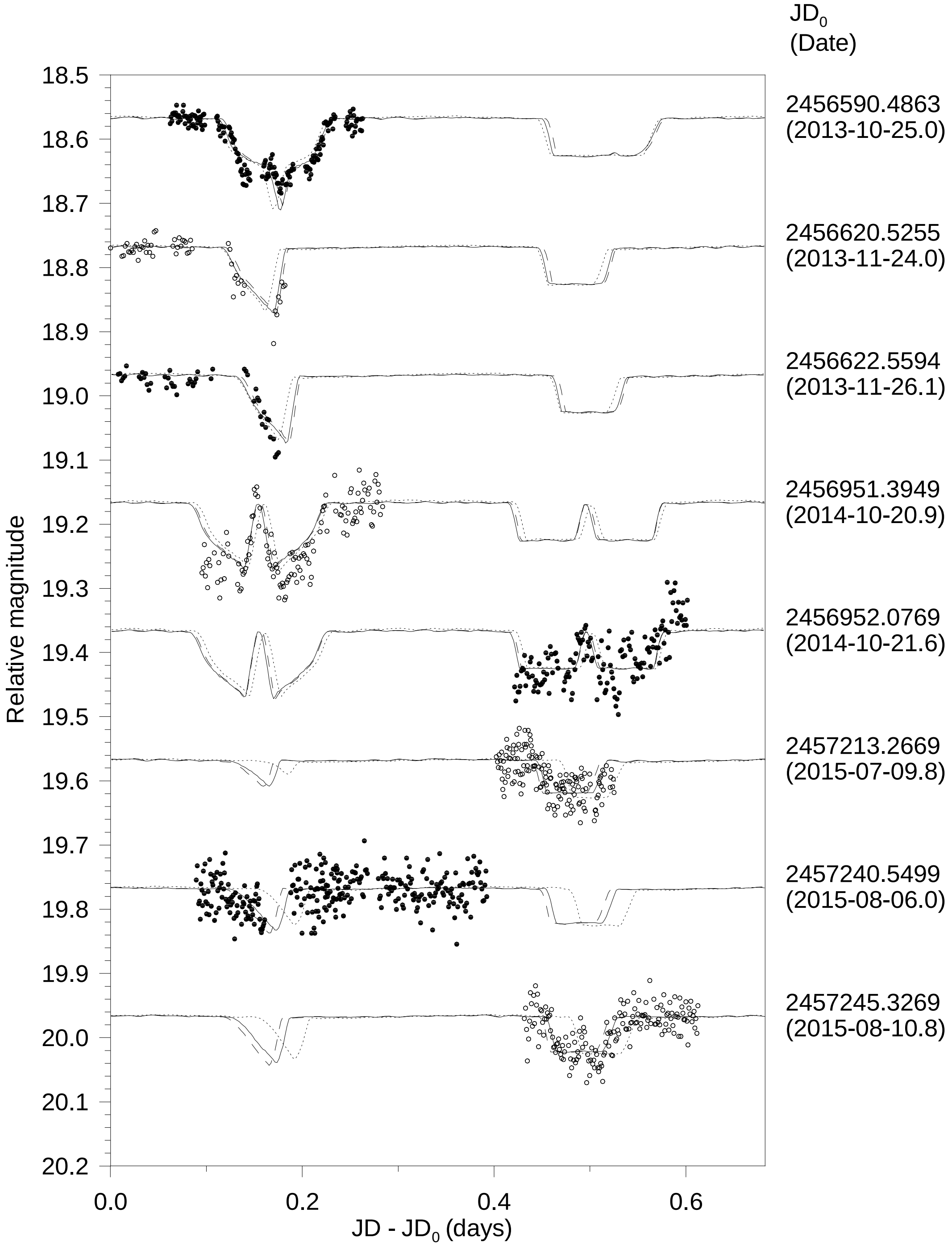

Fig. 10. Same as Fig. 9, but for data from later dates.

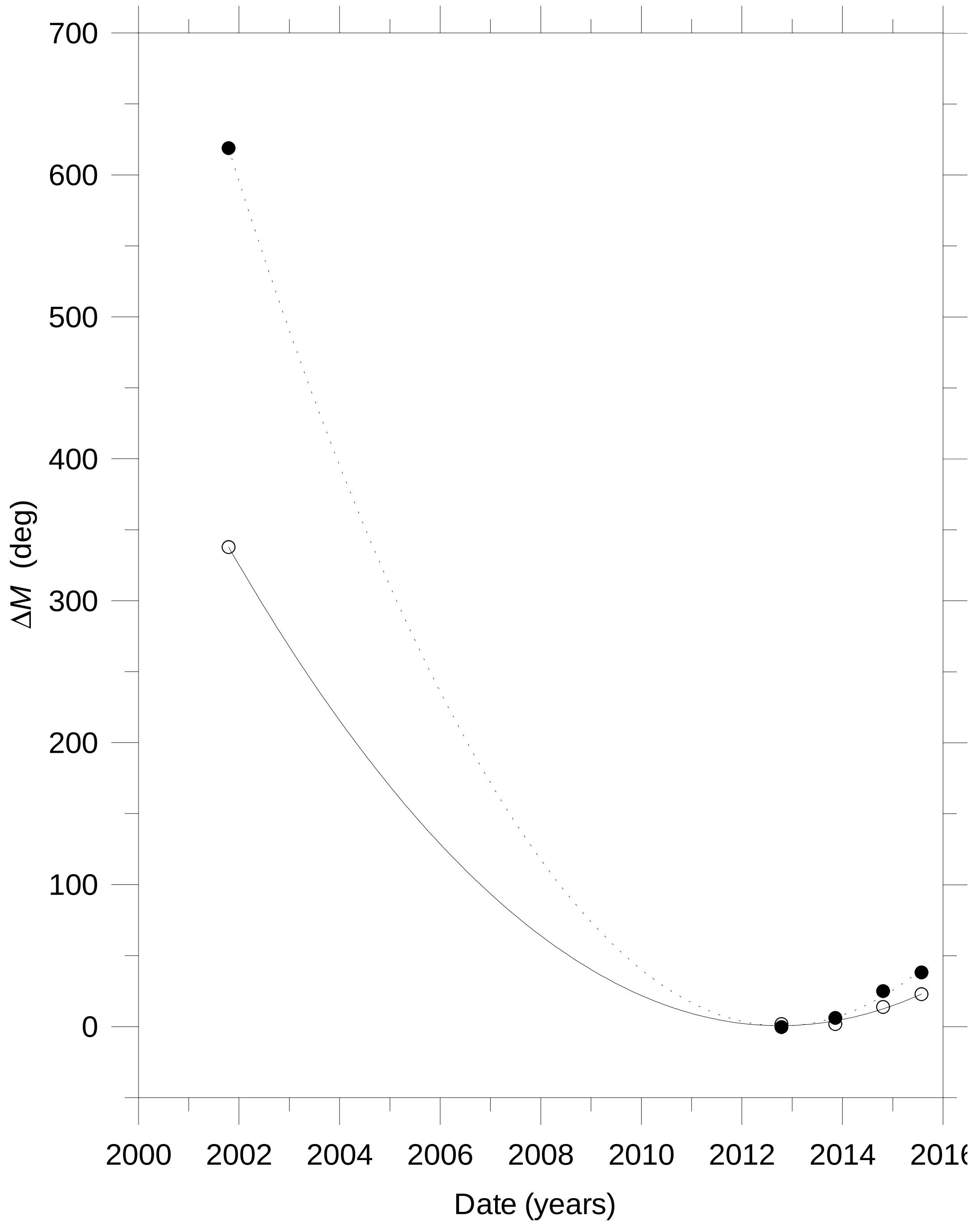

Fig. 11. Time evolutions of the mean anomaly difference $\Delta M$ for (88710) 2001 SL9. See text for details. Each point corresponds to the middle of one of the five apparitions from 2001 to 2015. The open and solid circles stand for the two solutions with $\Delta M_d = 2.8$ and 5.2 deg/yr$^2$, respectively. The sizes of the symbols in vertical direction represent estimated $3\sigma$ uncertainties in the timing of events ($\pm 5^\circ$ in mean anomaly). The curves are quadratic fits to the data points.

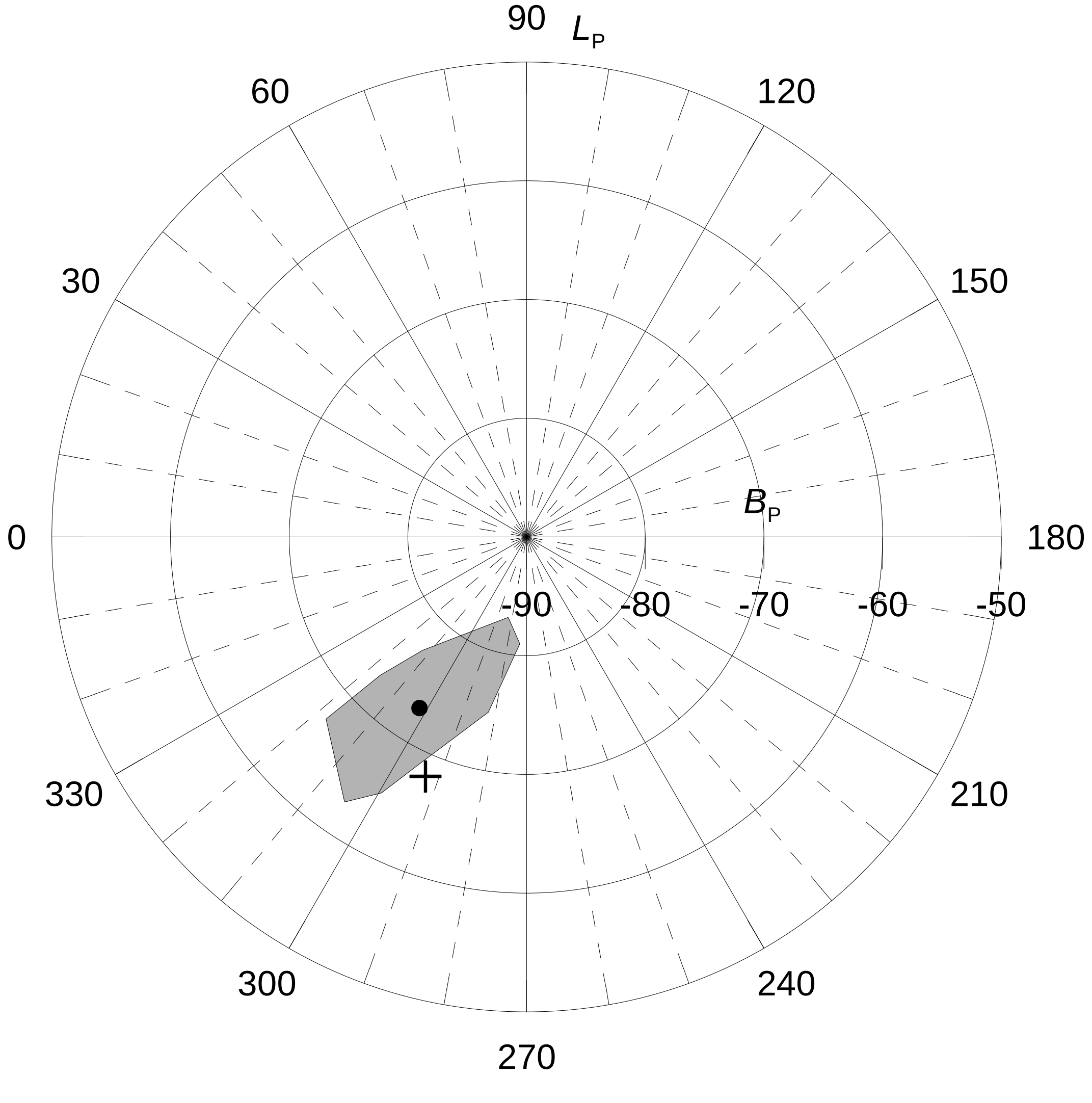

Fig. 12. Area of admissible poles for the mutual orbit of (88710) 2001 SL9 in ecliptic coordinates (grey area). The dot is the nominal solution given in Table 5. This area corresponds to $3\sigma$ confidence level. The south pole of the current asteroid's heliocentric orbit is marked with the cross.

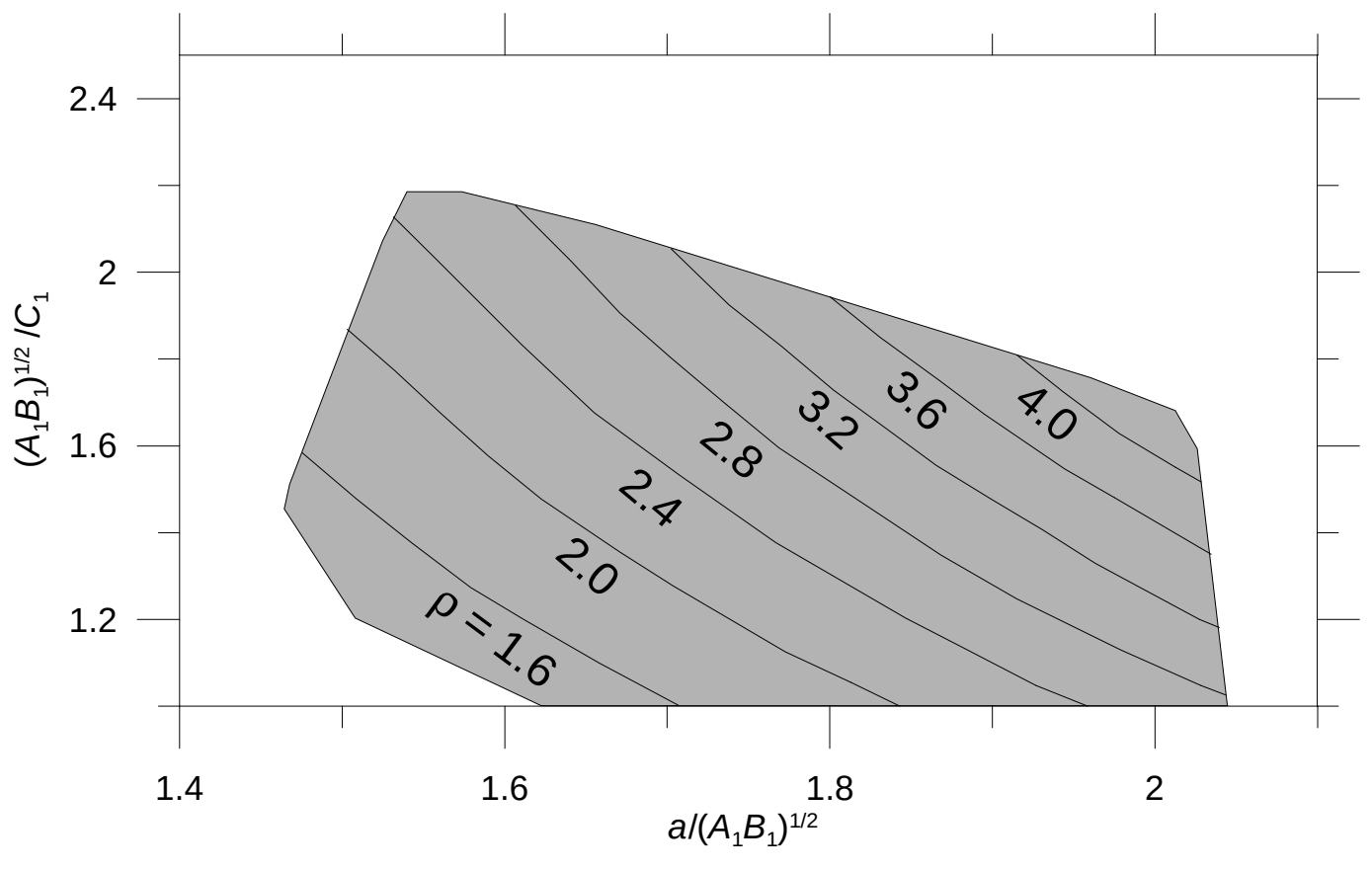

Fig. 13. Area of admissible combinations of the ratio between the mean equatorial and the polar axes of the primary ($(A_1B_1)^{1/2}/C_1$) and the semimajor axis of the mutual orbit $a$ of (88710) 2001 SL9. This area corresponds to $3\sigma$ confidence level. Values of the bulk density of the system ($\rho$) in g cm$^{-3}$ are indicated.

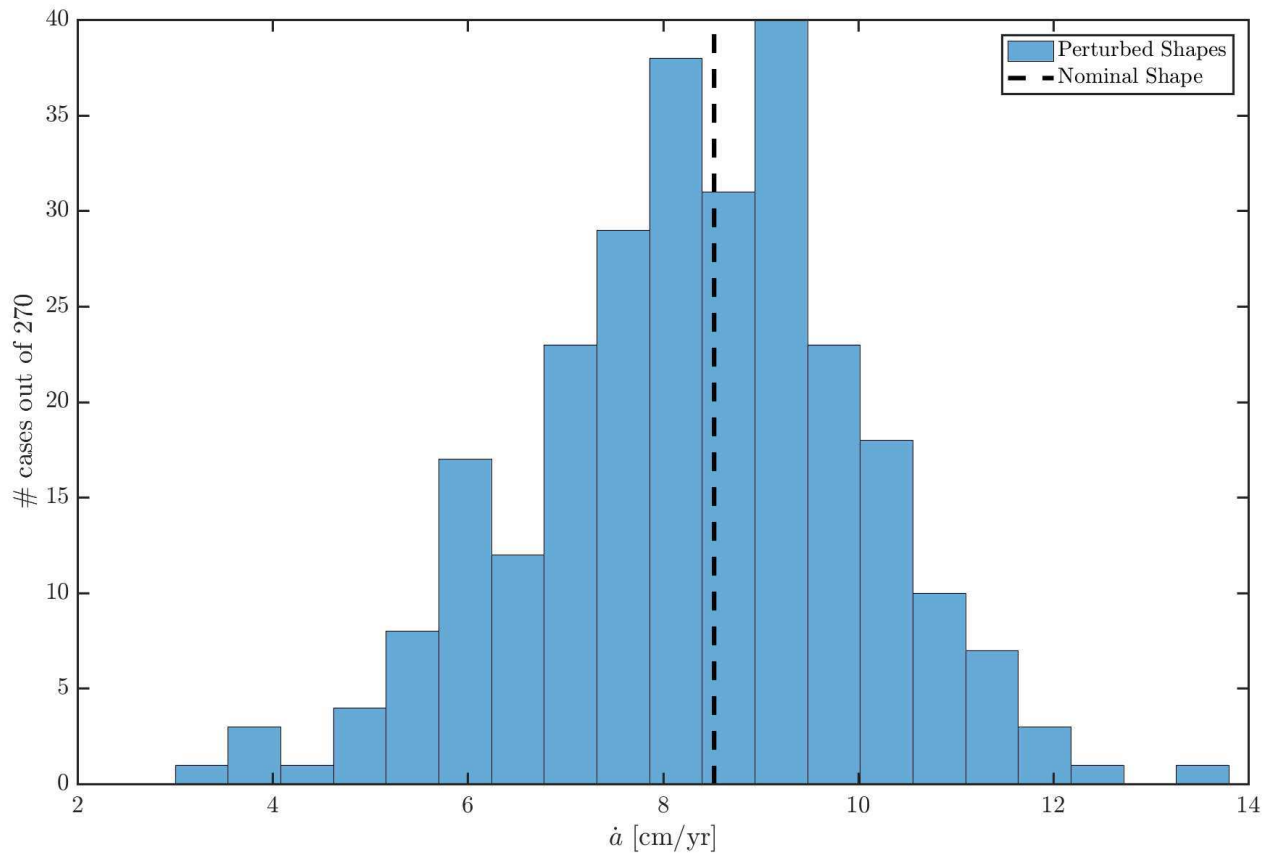

Fig. 14. Histogram of the resulting BYORP induced semimajor axis drift rates for the 270 perturbed secondary shape models of 1999 KW4.

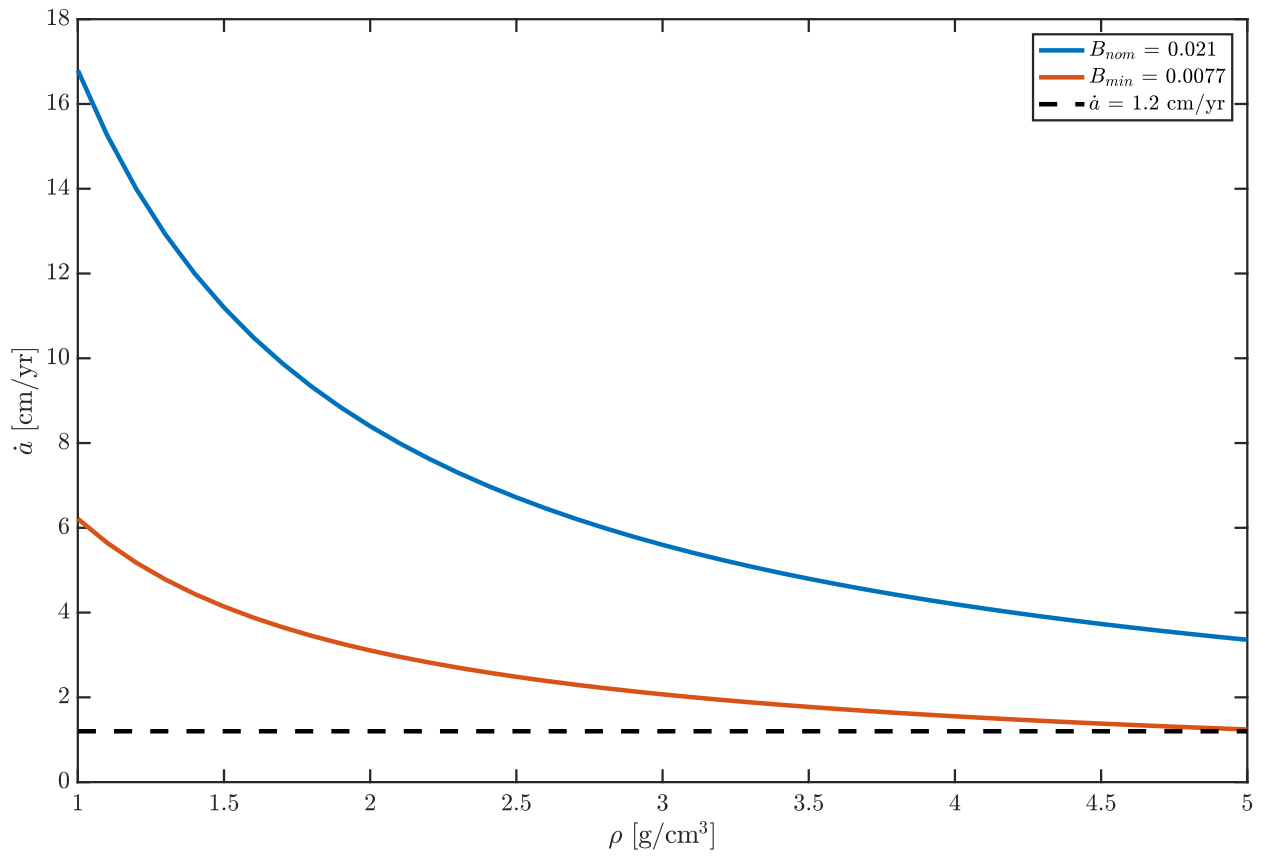

Fig. 15. Variation of $\dot{a}_B$ with secondary density, for the nominal and minimum BYORP coefficients.

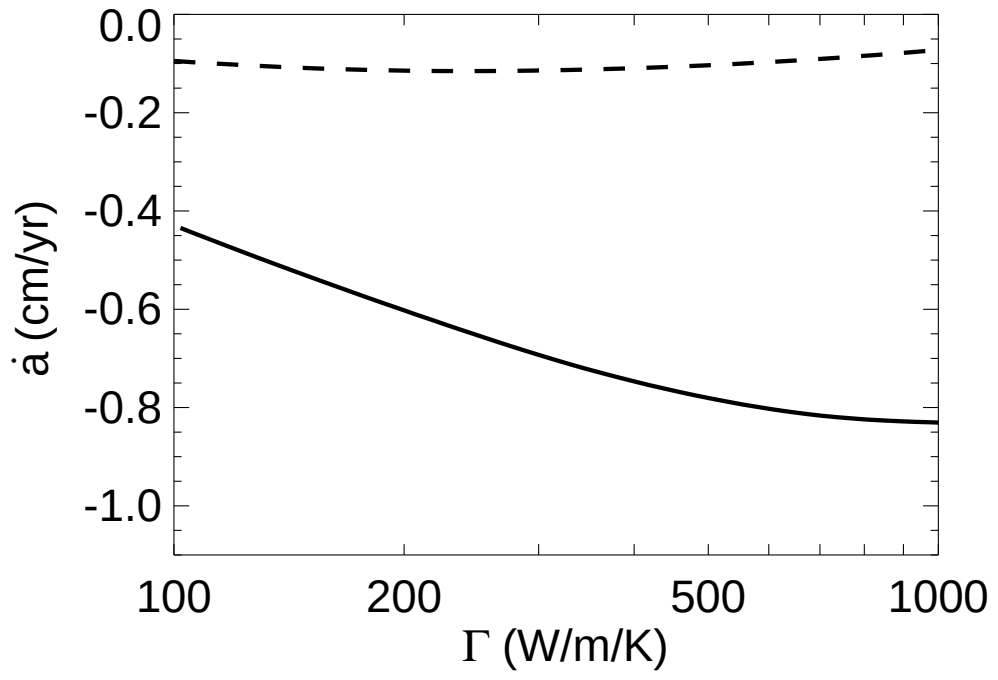

Fig. 16. Semimajor axis drift of the mutual orbit due to differential Yarkovsky effect as a function of thermal inetria Γ. Solid curve corresponds to (66391) 1999 KW4 and the dashed one corresponds to (88710) 2001 SL9.

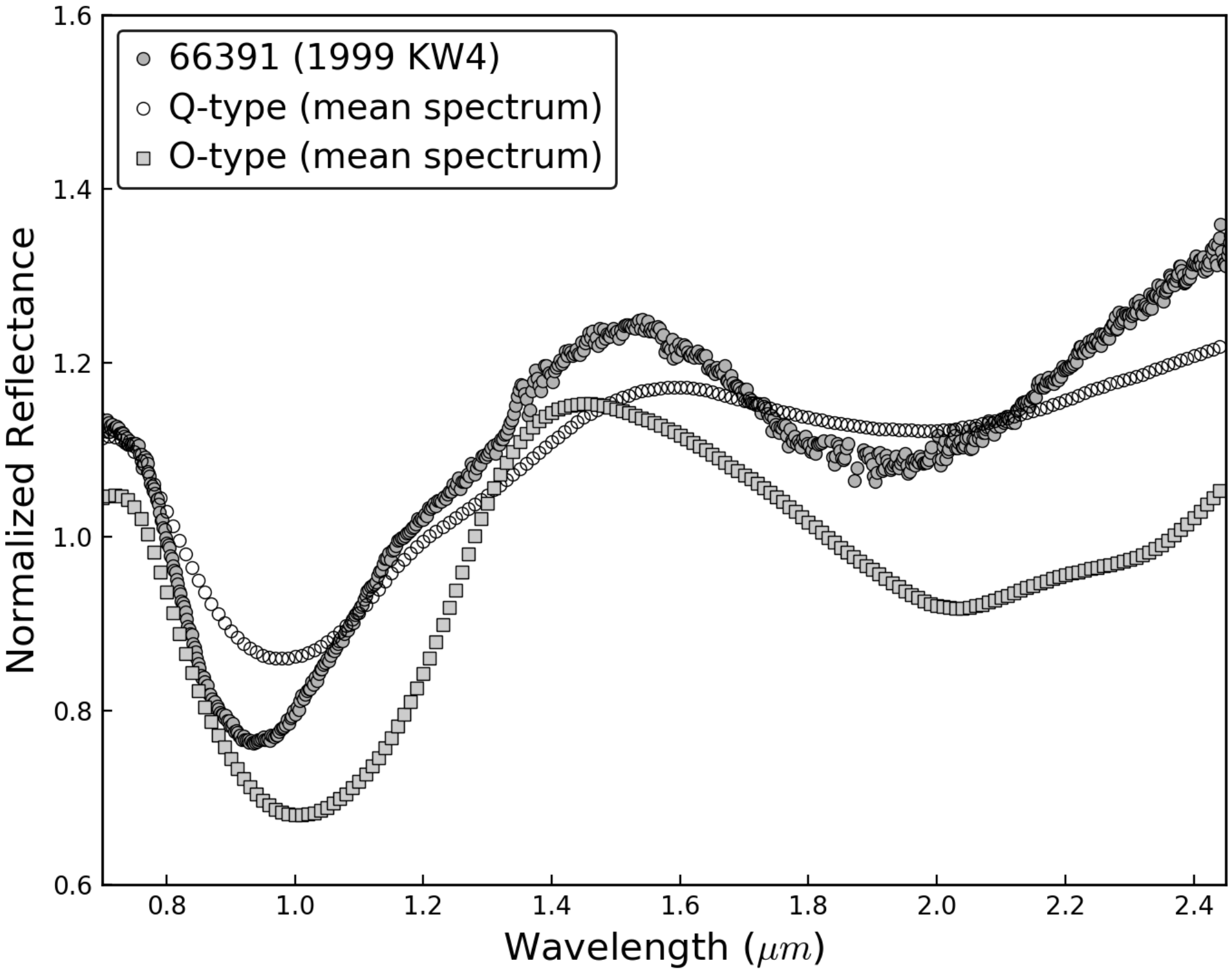

Fig. 17. NIR spectrum of (66391) 1999 KW4 and the mean spectrum of a Q-, and an O-type from DeMeo et al. (2009). All spectra are normalized to unity at 0.55 $\mu$m.